# Optimizing the Achievable Rate in MIMO Systems Assisted by Multiple Reconfigurable Intelligent Surfaces

N. Souto, *Senior Member, IEEE*, and João Carlos Silva, *Member, IEEE*

## Abstract

**In recent years there has been a growing interest in reconfigurable intelligent surfaces (RISs) as enablers for the realization of smart radio propagation environments which can provide performance improvements with low energy consumption in future wireless networks. However, to reap the potential gains of RIS it is crucial to jointly design both the transmit precoder and the phases of the RIS elements. Within this context, in this paper we study the use of multiple RIS panels in a parallel or multi-hop configuration with the aim of assisting a multi-stream multiple-input multiple-output (MIMO) communication. To solve the nonconvex joint optimization problem of the precoder and RIS elements targeted at maximizing the achievable rate, we propose an iterative algorithm based on the monotone accelerated proximal gradient (mAPG) method which includes an extrapolation step for improving the convergence speed and monitoring variables for ensuring sufficient descent of the algorithm. Based on the sufficient descent property we then present a detailed convergence analysis of the algorithm which includes expressions for the step size. Simulation results in different scenarios show that, besides being effective, the proposed approach can often achieve higher rates than other benchmarked schemes.**

This work is funded by FCT/MCTES through national funds and when applicable co-funded by EU funds under the project UIDB/50008/2020.

N. Souto and J. C. Silva are with the ISCTE-University Institute of Lisbon and Instituto de Telecomunicações, 1649-026 Lisboa, Portugal (e-mail: nuno.souto@lx.it.pt, joao.silva@iscte-iul.pt).

*Index Terms*—**Reconfigurable Intelligent Surface (RIS), achievable rate, Multiple-Input Multiple-Output (MIMO), proximal minimization.**

# I. INTRODUCTION

Over the last two decades we have been witnessing tremendous advances in wireless communication systems, which have enabled the support of a multitude of multimedia applications while also coping with the ever-growing number of mobile devices. Still, it is already clear that the full support of demanding applications such as immersive reality, holographic projection, autonomous driving, and tactile Internet, will only be feasible in future 6G and beyond networks [1]. Addressing this ever-increasing demand for higher data rates motivates further exploration of the mmWave bands that started to be integrated into 5G systems, and the move to even higher frequencies such as the Terahertz (THz) band [2]. While these bands can offer large available bandwidths they also pose several difficult challenges for implementing viable communication systems. One of the main problems lies in the very high propagation loss, which can drastically limit the signal coverage range. To address this limitation, one can resort to antenna arrays comprising a massive number of elements since the short wavelengths at these bands enable the deployment of elements with small footprints. Combined with multi-input multi-output (MIMO) beamforming techniques, these massive antenna arrays are able to provide high directional gains. However, the transmitted signals are still very susceptible to obstacles [3] which has raised the interest in recent years for the adoption of reconfigurable intelligent surfaces (RISs).

A RIS is an artificial planar structure, typically composed of a large number of low-cost, quasi-passive elements which can be independently tuned to induce a specific phase shift/amplitude on the reflected electromagnetic wave [4]. These elements can be reconfigured in real-time thus providing the capability to modify the wireless communication environment. Therefore, RISs have arisen has a potential solution for increasing coverage of communication systems as they have the capability to reflect and shape the incoming radio waves without the need for a power amplifier, RF chain, nor sophisticated signal processing [5][6]. This allows the use of RIS for accomplishing passive beamforming which can significantly improve the spectral efficiency with low energy consumption [7]-[9]. Still, realizing RIS-based communication systems is extremely challenging as recent design formulations have shown [10]. The optimization problems tend to be large, the RIS matrices and precoding matrices are coupled, and they typically include non-convex constraints such as constant modulus or discrete-valued phase shifts. Therefore, finding global optimal solutions efficiently is non-trivial. A strategy to simplify the problem is to independently design the precoder and RIS phases. This approach is followed within the context of RIS-aided MIMO THz systems in [11], where an adaptive gradient method is applied for optimizing the RIS phases. Once the RIS matrix is computed, the precoder and combiner matrices are then obtained using a singular value decomposition (SVD) of the cascaded channel. While the approach achieves large gains over a RIS with random phase, the complexity tends to become very high when working with large antenna arrays and RISs with many elements. Also relying on the independent optimization of the precoder and RIS, [12] presented a low complexity proximal gradient which was targeted at optimizing the RIS matrix and which resorts to simple element-wise normalization. Due to its simplicity, it can cope with large problem settings while providing large gains. Still, to obtain improved performance it is necessary to jointly optimize both the precoder and RIS. However, since finding the global optimal solution is difficult, most works try to propose efficient approaches for finding local optimal solutions. For example, the work in [13] considered a joint beamforming scheme based on fractional programming optimization for RIS-assisted THz communications. Even though it corresponds to a limited scenario, it showed encouraging results. In [14] the authors targeted the minimization of the transmitted power constrained on the signal to interference plus noise ratio required at the receivers. To solve the problem, they proposed an alternating optimization (AO) algorithm that iteratively optimizes the RIS phase shifts and the transmit beamforming vector. The coverage and achievable rate performance are significantly improved when compared to conventional systems without RIS. However, it requires many iterations, especially when the number of RIS elements is large. Another iterative algorithm based on the AO approach was proposed in [15] but it only optimizes the transmit covariance matrix or one of the reflection coefficients with all the other variables being fixed. While this allows the implementation complexity to be reduced, it can still be high when the number of reflecting elements is large. In [16], the authors addressed the joint optimization of the transmit beamformer and RIS phase shifts with the objective of maximizing the achievable rate in single-antenna single-user scenarios. Algorithms based on the fixed-point iteration method and manifold optimization were proposed for solving the problem. Also aiming at maximizing the achievable rate but in a multi-stream MIMO system, the authors in [17] proposed an iterative projected gradient method (PGM) for solving the joint optimization problem of the covariance matrix of the transmitted signal and the RIS elements. The PGM algorithm was shown to achieve similar achievable rates to the AO method from [14] in single-user scenarios but requiring with fewer iterations.

While all the previous works consider single RIS scenarios, the adoption of multiple RIS could enable additional degrees of freedom and potentially ensure uninterrupted connectivity inside a target area, as discussed in [18]. Therefore, a few recent works with multiple RIS have started to appear [19]-[24]. Within this scope, the work in [21] studied the joint active and passive precoding design, where multiple parallel RISs assist a single stream transmission between a base station (BS) and a single-antenna receiver. In this case, a semidefinite relaxation-based method was proposed for solving the problem. Also concerning multiple parallel RIS, the work in [22] considered a simpler approach where the RIS matrices are first optimized followed by the computation of the precoder/combiner using the SVD. Regarding the use of multiple RISs in a multi-hop configuration, a few recent studies have started to emerge such as in [23], which emphasized the need for innovative design approaches and optimization methods in this type of systems. To cope with this complicate scenario, the authors in [24] applied machine learning methods, in particular a deep reinforcement learning based framework, to find feasible solutions to the design problem of transmit beamforming at the BS and phase shift matrices at the RISs.

Motivated by the work above, in this paper we study the realization of smart radio propagation environments where multiple passive RIS panels deployed in the vicinity of a BS assist a transmission towards a user. We consider a multi-stream MIMO configuration and address the joint design problem of the precoder and of all the RIS elements in the panels targeted at maximizing the achievable rate. To tackle this nonconvex problem, we apply the monotone accelerated proximal gradient (mAPG) method [25] in order to obtain an iterative algorithm that is simple to implement and is guaranteed to converge to a critical point of the problem. The mAPG method includes an extrapolation step based on a linear combination of the previous and current solution estimates with the aim of improving the convergence speed which is typically slow in gradient-based methods. It also includes the computation of additional monitoring variables for ensuring sufficient descent of the algorithm. This property facilitates the convergence analysis of the algorithm which, otherwise would be difficult to perform due to the nonconvex nature of the problem. The mAPG approach is first developed for multiple parallel RISs and then extended to multi-hop scenarios.

The main contributions of the paper can be summarized as follows:

- We formulate the joint achievable rate maximization problem defined over the precoder and the phase shifts of all RIS elements, considering a multi-stream MIMO communication system assisted by multiple parallel RIS panels. To find a solution for this problem we derive an mAPG-based iterative algorithm, presenting exact expressions for the required gradients and proximal operators.
- We show that the objective function in the optimization problem has Lipschitz continuous gradients and provide a bound for the Lipschitz constants. Using the Lipschitz continuity of the gradients we then prove that the proposed joint precoding and RIS mAPG based (JPR-mAPG) algorithm is guaranteed to converge to a critical point.
- We extend the JPR-mAPG algorithm for multi-hop RIS-assisted scenarios, providing expressions for the required gradients. We establish the Lipschitz continuity of the gradients also for this scheme, presenting the respective bound for the Lipschitz constants. This ensures that the convergence of the algorithm is still valid for the multi-hop case.
- The achievable rate of the proposed approach is evaluated through numerical simulations and compared against the R-APG algorithm that only optimizes the RIS phases shifts, and against the PGM algorithm which, while being a gradient-based algorithm, does not apply acceleration. Results show that, while the adoption of some type of optimization in the design of the RIS can offer clear gains over a static approach, the proposed JPR-mAPG approach is often able to provide higher rates. Furthermore, the use of several panels can have a substantial impact on the achieved rates, but the improvements have a strong dependency on the location of the panels. In the case of multi-hop, it was shown that with multiple panels it is possible to provide signal coverage when severe blockage between transmitter and receiver exists.

The rest of the paper is organized as follows. Section II defines de system model and optimization problem for the multiple parallel RIS-assisted scheme, presenting the proposed JPR-mAPG algorithm and providing the respective convergence analysis. Section III extends the work to a multi-hop RIS-assisted scheme. Section IV presents several simulation results in different scenarios, whereas the conclusions are provided in Section V.

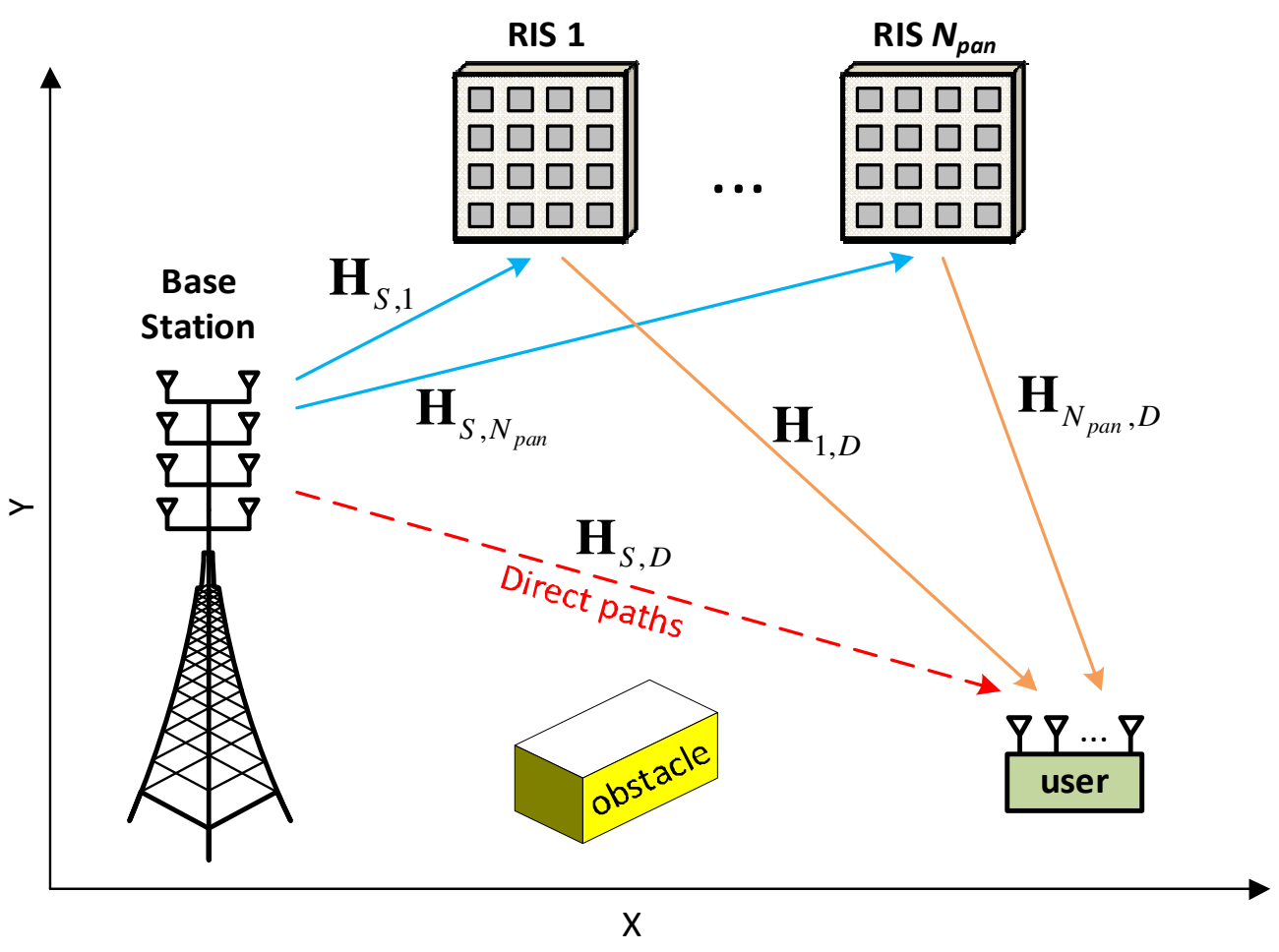

Fig. 1. System model layout considered for the multiple parallel RIS-assisted MIMO communication system.

*Notation:* Matrices and vectors are represented by uppercase and lowercase boldface letters, respectively. $(\cdot)^T$ and $(\cdot)^H$ denote the transpose and conjugate transpose of a matrix/vector, $\|\cdot\|_2$ is the 2-norm of a vector/matrix, $\|\cdot\|_F$ is the Frobenius norm, diag$(\cdot)$ represents a diagonal of a matrix if the argument is a matrix or represents a diagonal matrix if the argument is a vector, $\underset{i=1,\ldots N}{\text{blkdiag}}\{\mathbf{A}_i\}$ is a block diagonal matrix whose diagonal elements are matrices $\mathbf{A}_i$ ($i=1\ldots N$), $\partial g(\mathbf{x})$ is the sub-differential set of $g(\mathbf{x})$, $\mathbf{I}_n$ is the $n\times n$ identity matrix and $\mathcal{I}_{\mathcal{D}}(\mathbf{v})$ is the indicator function which returns 0 if $\mathbf{v}\in\mathcal{D}$ and $+\infty$ otherwise. We also consider that for matrices we have $\prod_{i=1}^{n}\mathbf{A}_i=\mathbf{A}_n\times\ldots\times\mathbf{A}_1$.

## II. Multiple parallel RIS-assisted communication

### A. *System Model and Problem Formulation*

First, we address the case where, to help establish the communication, the receiver is served by $N_{PAN}$ RIS panels simultaneously, as illustrated in Fig. 1. The transmitted signal comprises $N_s$ data streams and is represented as $\mathbf{s}=\left[s_1\ldots s_{N_s}\right]^T$, where $s_i\in\mathbb{C}$ corresponds to an amplitude and phase modulated symbol with $\mathrm{E}\left[\|\mathbf{s}\|^2\right]=N_s$. The signal arriving at the receiver array, can be written as

$$\mathbf{r}=\sqrt{\rho}\mathbf{HFs}+\mathbf{n} \tag{1}$$

where $\sqrt{\rho}$ denotes the power per stream, $\mathbf{H}\in\mathbb{C}^{N_{rx}\times N_{tx}}$ is the RIS dependent channel matrix,

$\mathbf{F} \in \mathbb{C}^{N_{tx} \times N_s}$ is the BS precoder matrix and $\mathbf{n} \in \mathbb{C}^{N_{rx} \times 1}$ is the noise vector, which contains independent zero-mean circularly symmetric Gaussian samples with covariance $\sigma_n^2 \mathbf{I}_{N_{rx}}$.

We consider the existence of blockages between multiple RIS which, combined with the high path-loss at high frequencies, will tend to result in negligible powers in the signals that are reflected by the RISs two or more times. Therefore, similarly to [22], we ignore the cross-reflections and approximate the channel matrix as

$$\mathbf{H} = \mathbf{H}_{S,D} + \sum_{i=1}^{N_{pan}} \mathbf{H}_{i,D} \mathbf{\Phi}_i \mathbf{H}_{S,i} \ , \tag{2}$$

where $\mathbf{H}_{S,D} \in \mathbb{C}^{N_{rx} \times N_{tx}}$ denotes the direct channel between the transmitter and receiver, $\mathbf{H}_{S,i} \in \mathbb{C}^{N_{ris} \times N_{tx}}$ is the channel matrix between the transmitter and the $i^{\text{th}}$ RIS panel, and $\mathbf{H}_{i,D} \in \mathbb{C}^{N_{rx} \times N_{ris}}$ is the channel matrix between the $i^{\text{th}}$ RIS panel and the receiver. Matrix $\mathbf{\Phi}_i \in \mathbb{C}^{N_{ris} \times N_{ris}}$ models the phase shifting effect of the $i^{\text{th}}$ RIS panel. This matrix has a diagonal structure and can be written as $\mathbf{\Phi}_i = \text{diag}(\boldsymbol{\varphi}_i)$ with $\boldsymbol{\varphi}_i = \left[ a e^{j\theta_1^i}, ..., a e^{j\theta_{N_{ris}}^i} \right]^T$, where $\theta_m^i \in [0, 2\pi)$ ( $m = 1...N_{pan}$ ) represents the phase shift of the $m^{\text{th}}$ RIS element and $a$ denotes the amplitude of the reflection coefficients. Along the paper we will assume perfect channel state information (CSI) at the BS. Even though, obtaining CSI in RIS aided schemes is challenging, several methods have been proposed in recent years [10][26].

In this paper we design the precoder and RIS matrices to maximize the achievable rate of the system which, assuming Gaussian signaling, can be written as [27]

$$R = \log_2 \ \det \left( \mathbf{I}_{N_s} + \frac{\rho}{P_n} \mathbf{F}^H \mathbf{H}^H \mathbf{H} \mathbf{F} \right) \tag{3}$$

in bits/s/Hz, with $P_n$ denoting the noise power (i.e., $P_n = \sigma_n^2$ ).It is important to remind that the channel matrix $\mathbf{H}$, and thus also the rate, depend on the RIS coefficient vectors $\boldsymbol{\varphi}_i$, $i = 1...N_{pan}$.

The optimization problem can then be formulated as

$$\min_{\mathbf{F}, \boldsymbol{\varphi}_i, \ i=1...N_{RIS}} \ f\left(\mathbf{F}, \boldsymbol{\varphi}_1, ..., \boldsymbol{\varphi}_{N_{pan}}\right) = \text{-} \ln \det \left( \mathbf{I}_{N_s} + \frac{\rho}{P_n} \mathbf{F}^H \mathbf{H}^H \mathbf{H} \mathbf{F} \right) \tag{4a}$$

$$\text{subject to} \quad \|\mathbf{F}\|_F^2 \leq N_s, \tag{4b}$$

$$\boldsymbol{\varphi}_i \in \mathcal{U}_{N_{RIS}}, \ i = 1...N_{pan}. \tag{4c}$$

where $\mathcal{U}_{N_{RIS}} = \left\{ \mathbf{v} \in \mathbb{C}^{N_{RIS} \times 1} : |v_m| = a, \ m = 1...N_{RIS} \right\}$ is the set of all complex valued vectors whose elements have a magnitude of $a$.

### B. Accelerated Proximal Gradient based Solution

The coupling between optimization variables $\mathbf{F}$ and $\boldsymbol{\varphi}_i$ in (4a), as well as constraint (4c) make the problem nonconvex and difficult to solve effciently. Furthermore, when considering a typical RIS assisted massive or ultra-massive MIMO scenario, formulation (4) will typically correspond to a large-scale optimization problem. This motivates the resort to first-order methods with low iteration cost for solving the problem. In this case, we propose an algorithm based on the application of the monotone Accelerated Proximal Gradient (mAPG) method for nonconvex optimization problems described in [25]. First, we use the indicator function to encode constraints (4b) and (4c) and rewrite problem (4) as

$$\min_{\mathbf{F},\boldsymbol{\varphi}_i,\, i=1\ldots N_{RIS}} f\left(\mathbf{F},\boldsymbol{\varphi}_1,\ldots,\boldsymbol{\varphi}_{N_{pan}}\right)+\mathcal{I}_{\mathcal{C}}\left(\mathbf{F}\right)+\sum_{i=1}^{N_{pan}}\mathcal{I}_{\mathcal{U}_{N_{RIS}}}\left(\boldsymbol{\varphi}_i\right) \tag{5}$$

where $\mathcal{C}=\left\{\mathbf{X}\in\mathbb{C}^{N_{tx}\times N_s}:\left\|\mathbf{F}\right\|_F^2\leq N_s\right\}$. To find a solution for (5) we apply a gradient step consisting of $\mathbf{F}^{(q)}-\alpha\nabla_{\mathbf{F}^*}f\left(\mathbf{F}^{(q)}\right)$ and $\boldsymbol{\varphi}_i^{(q)}-\alpha\nabla_{\boldsymbol{\varphi}^*}f\left(\boldsymbol{\varphi}_i^{(q)}\right)$, followed by a proximal mapping, whose operator for a function $g$ is defined as

$$\operatorname{prox}_g\left(\mathbf{a}\right)=\underset{\hat{\mathbf{x}}}{\operatorname{argmin}}\;\; g\left(\hat{\mathbf{x}}\right)+\frac{1}{2}\left\|\hat{\mathbf{x}}-\mathbf{a}\right\|^2. \tag{6}$$

To improve the typical slow convergence rate of gradient-based methods, we can apply an extrapolation step as proposed in [28], which involves the computation of additional variables $\left(\mathbf{P}^{(q)},\mathbf{y}_1^{(q)},\ldots,\mathbf{y}_{N_{pan}}^{(q)}\right)$. However, this extrapolation may result in a bad update which, while not causing the objective function to increase, makes the convergence difficult to analyze, especially due to the nonconvex nature of the problem. Therefore, we add an extra set of variables $\left(\mathbf{U}^{(q)},\mathbf{v}_1^{(q)},\ldots,\mathbf{v}_{N_{pan}}^{(q)}\right)$ whose role is to monitor and correct the update of $\left(\mathbf{F}^{(q)},\boldsymbol{\varphi}_1^{(q)},\ldots,\boldsymbol{\varphi}_{N_{pan}}^{(q)}\right)$. Algorithm 1 summarizes all the steps of the proposed approach. Variables $\alpha$, $Q$ and $t_k$ denote the step size, maximum number of iterations and extrapolation parameter, respectively. In step 5 and 6, the proximal operator of the indicator functions corresponds to the Euclidean projection over the respective sets and can be computed as

**Algorithm 1:** Joint Precoding and RIS - Monotone Accelerated Proximal Gradient based Algorithm (JPR-mAPG)

**Input:** $\mathbf{r}$, $\mathbf{H}$, $\alpha$, $Q$, $\mathbf{F}^{(0)}$, $\boldsymbol{\varphi}_i^{(0)}$

$\hat{\mathbf{x}}^{(0)} = \mathbf{z}^{(0)} = 0$, $t_0 = 0$, $t_1 = 1$

**for** $q$=1,1,…$Q$ **do**

$$\mathbf{P}^{(q)} = \mathbf{F}^{(q)} + \frac{t_{q-1}}{t_q}\left(\mathbf{W}^{(q)} - \mathbf{F}^{(q)}\right) + \frac{t_{q-1}-1}{t_q}\left(\mathbf{F}^{(q)} - \mathbf{F}^{(q-1)}\right)$$

$$\mathbf{y}_i^{(q)} = \boldsymbol{\varphi}_i^{(q)} + \frac{t_{q-1}}{t_q}\left(\mathbf{z}^{(q)} - \boldsymbol{\varphi}_i^{(q)}\right) + \frac{t_{q-1}-1}{t_q}\left(\boldsymbol{\varphi}_i^{(q)} - \boldsymbol{\varphi}_i^{(q-1)}\right), \quad i = 1,\dots,N_{pan}$$

$$\mathbf{W}^{(q+1)} = \mathrm{prox}_{\alpha\mathcal{I}_{\mathcal{C}}}\left(\mathbf{P}^{(q)} - \alpha\nabla_{\mathbf{F}^*} f\left(\mathbf{P}^{(q)}\right)\right)$$

$$\mathbf{z}_i^{(q+1)} = \mathrm{prox}_{\alpha\mathcal{I}_{\mathcal{U}_{N_{RIS}}}}\left(\mathbf{y}_i^{(q)} - \alpha\nabla_{\boldsymbol{\varphi}_i^*} f\left(\mathbf{y}_i^{(q)}\right)\right), \qquad i = 1,\dots,N_{pan}$$

$$\mathbf{U}^{(q+1)} = \mathrm{prox}_{\alpha\mathcal{I}_{\mathcal{C}}}\left(\mathbf{F}^{(q)} - \alpha\nabla_{\mathbf{F}^*} f\left(\mathbf{F}^{(q)}\right)\right)$$

$$\mathbf{v}_i^{(q+1)} = \mathrm{prox}_{\alpha\mathcal{I}_{\mathcal{U}_{N_{RIS}}}}\left(\boldsymbol{\varphi}_i^{(q)} - \alpha\nabla_{\boldsymbol{\varphi}_i^*} f\left(\boldsymbol{\varphi}_i^{(q)}\right)\right), \qquad i = 1,\dots,N_{pan}$$

**if** $f\left(\mathbf{W}^{(q+1)}, \mathbf{z}_1^{(q+1)}, \dots, \mathbf{z}_{N_{pan}}^{(q+1)}\right) \le f\left(\mathbf{U}^{(q+1)}, \mathbf{v}_1^{(q+1)}, \dots, \mathbf{v}_{N_{pan}}^{(q+1)}\right)$ **then**

$\mathbf{F}^{(q+1)} = \mathbf{W}^{(q+1)}$

$\boldsymbol{\varphi}_i^{(q+1)} = \mathbf{z}_i^{(q+1)}, \quad i = 1,\dots,N_{pan}$

**else**

$\mathbf{F}^{(q+1)} = \mathbf{U}^{(q+1)}$

$\boldsymbol{\varphi}_i^{(q+1)} = \mathbf{v}_i^{(q+1)}, \quad i = 1,\dots,N_{pan}$

**end if**

$$t_{q+1} = \frac{1}{2}\left(\sqrt{4\left(t_q\right)^2 + 1} + 1\right)$$

**end for**

**Output:** $\mathbf{F}^{(q+1)}, \boldsymbol{\varphi}_i^{(q+1)}, \quad i = 1,\dots,N_{pan}$.

$$\mathrm{prox}_{\alpha\mathcal{I}_{\mathcal{C}}}\left(\mathbf{X}\right) = \begin{cases} \mathbf{X}, & \left\|\mathbf{X}\right\|_F^{\,2} \le N_s \\ \dfrac{\sqrt{N_s}}{\left\|\mathbf{X}\right\|_F}\mathbf{X}, & \text{otherwise} \end{cases} \tag{7}$$

and

$$\mathrm{prox}_{\alpha\mathcal{I}_{\mathcal{U}_{N_{RIS}}}}\left(\mathbf{x}\right) = a\left(\mathbf{x} \oslash \left|\mathbf{x}\right|\right) , \tag{8}$$

where $\oslash$ denotes the Hadamard division and $\left|\cdot\right|$ represents the absolute value applied elementwise. Regarding the later projection, we note that in the special case of $\mathrm{prox}_{\alpha\mathcal{I}_{\mathcal{U}_1}}\left(0\right)$, any point of the form $ae^{j\theta}$ with $\theta \in \left[0, 2\pi\right)$ is a valid solution. The complex-valued gradient of $f\left(\mathbf{F}, \boldsymbol{\varphi}_1, \dots, \boldsymbol{\varphi}_{N_{pan}}\right)$ required in the proposed algorithm can be computed according to the following Lemma.

*Lemma 1:* Let $f\left(\mathbf{F},\boldsymbol{\varphi}_1,...,\boldsymbol{\varphi}_{N_{pan}}\right)$ be defined as in (4a) with the channel matrix represented as (2). Then, the corresponding complex-valued gradient with respect to $\mathbf{F}^*$ and $\boldsymbol{\varphi}_i^*$ ($i=1,...,N_{pan}$) can be computed as

$$\nabla_{\mathbf{F}^*} f\left(\mathbf{F},\boldsymbol{\varphi}_1,...,\boldsymbol{\varphi}_{N_{pan}}\right) = -\frac{\rho}{P_n}\mathbf{H}^H\mathbf{H}\mathbf{F}\left(\mathbf{I}_{N_s}+\frac{\rho}{P_n}\mathbf{F}^H\mathbf{H}^H\mathbf{H}\mathbf{F}\right)^{-1} \tag{9}$$

$$\nabla_{\boldsymbol{\varphi}_i^*} f\left(\mathbf{F},\boldsymbol{\varphi}_1,...,\boldsymbol{\varphi}_{N_{pan}}\right) = -\frac{\rho}{P_n}\operatorname{diag}\left[{\mathbf{H}_{i,D}}^H\mathbf{H}\mathbf{F}\left(\mathbf{I}_{N_s}+\frac{\rho}{P_n}\mathbf{F}^H\mathbf{H}^H\mathbf{H}\mathbf{F}\right)^{-1}\mathbf{F}^H{\mathbf{H}_{S,i}}^H\right] \tag{10}$$

*Proof:* Proof is shown in Appendix A. ■

## *C. Convergence Analysis*

In this subsection we show that with the proper selection of the step size, the algorithm can be guaranteed to converge to a critical point. To help prove the convergence of the algorithm we first present the following theorem.

*Theorem 1:* Function $f\left(\mathbf{F},\boldsymbol{\varphi}_1,...,\boldsymbol{\varphi}_{N_{pan}}\right)$ defined as in (4a) with the channel matrix represented as (2) has Lipschitz continuous gradients over feasible set (4b)-(4b), i.e.,

$$\begin{aligned}&\Bigg(\left\|\nabla_{\mathbf{F}^*} f\left(\mathbf{F}^{(q_2)},\boldsymbol{\varphi}_1^{(q_2)},...,\boldsymbol{\varphi}_{N_{pan}}^{(q_2)}\right)-\nabla_{\mathbf{F}^*} f\left(\mathbf{F}^{(q_1)},\boldsymbol{\varphi}_1^{(q_1)},...,\boldsymbol{\varphi}_{N_{pan}}^{(q_1)}\right)\right\|_F^2+\\&+\sum_{i=1}^{N_{pan}}\left\|\nabla_{\tilde{\boldsymbol{\varphi}}_i^*} f\left(\mathbf{F}^{(q_2)},\boldsymbol{\varphi}_1^{(q_2)},...,\boldsymbol{\varphi}_{N_{pan}}^{(q_2)}\right)-\nabla_{\tilde{\boldsymbol{\varphi}}_i^*} f\left(\mathbf{F}^{(q_1)},\boldsymbol{\varphi}_1^{(q_1)},...,\boldsymbol{\varphi}_{N_{pan}}^{(q_1)}\right)\right\|_2^2\Bigg)^{\frac{1}{2}}\\&\le L\left(\left\|\mathbf{F}^{(q_2)}-\mathbf{F}^{(q_1)}\right\|_F^2+\sum_{i=1}^{N_{pan}}\left\|\boldsymbol{\varphi}_i^{(q_2)}-\boldsymbol{\varphi}_i^{(q_1)}\right\|_2^2\right)^{\frac{1}{2}},\end{aligned} \tag{11}$$

where constant $L$ is given by

$$L=\sqrt{\max\left(b^2+bc+d^2+de,c^2+bc+e^2+de\right)}, \tag{12}$$

with

$$b=\frac{\rho}{P_n}\zeta^2\left(1+2\frac{N_s\rho}{P_n}\zeta^2\right) \tag{13}$$

$$c=2\sqrt{N_s}\frac{\rho}{P_n}\zeta\lambda_{\max}\left(\tilde{\mathbf{H}}_{R,D}\right)\lambda_{\max}\left(\tilde{\mathbf{H}}_{S,R}\right)\left(1+\frac{N_s\rho}{P_n}\zeta^2\right) \tag{14}$$

$$d=2\sqrt{N_s}\frac{\rho}{P_n}\zeta\left(1+\frac{N_s\rho}{P_n}\zeta^2\right)\sum_{i=1}^{N_{pan}}\lambda_{\max}\left(\mathbf{H}_{i,D}\right)\lambda_{\max}\left(\mathbf{H}_{S,i}\right) \tag{15}$$

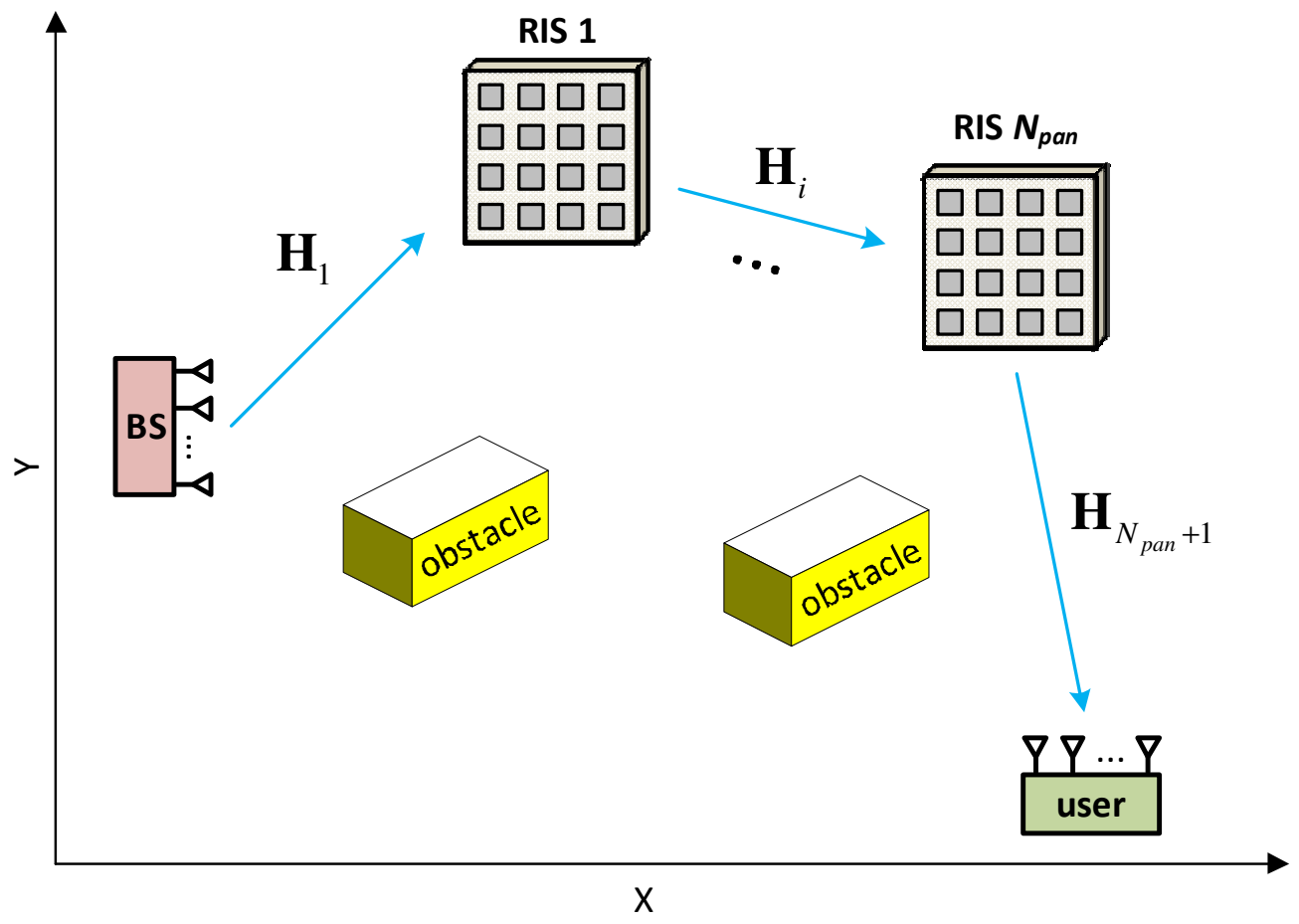

Fig. 2. System model layout considered for the multi-hop RIS-assisted MIMO communication system.

$$e = \frac{N_s\rho}{P_n}\left(1+2\frac{N_s\rho}{P_n}\zeta^2\right)\lambda_{\max}\left(\tilde{\mathbf{H}}_{R,D}\right)\lambda_{\max}\left(\tilde{\mathbf{H}}_{S,R}\right)\sum_{i=1}^{N_{pan}}\lambda_{\max}\left(\mathbf{H}_{i,D}\right)\lambda_{\max}\left(\mathbf{H}_{S,i}\right) \tag{16}$$

$$\zeta = \lambda_{\max}\left(\mathbf{H}_{S,D}\right)+a\sum_{i=1}^{N_{pan}}\lambda_{\max}\left(\mathbf{H}_{i,D}\right)\lambda_{\max}\left(\mathbf{H}_{S,i}\right) \tag{17}$$

and $\lambda_{\max}\left(\mathbf{X}\right)$ denoting the largest singular value of matrix $\mathbf{X}$.

*Proof:* Proof is shown in Appendix B. ■

After establishing the Lipschitz continuity of the gradients, the convergence of the algorithm can be stated in the following theorem.

*Theorem 2:* Let the step size in Algorithm 1 be selected as $\alpha < \frac{1}{L}$. Then both sequence sets $\left\{\left(\mathbf{F}^{(q)},\boldsymbol{\varphi}_1^{(q)},\ldots,\boldsymbol{\varphi}_{N_{pan}}^{(q)}\right)\right\}$ and $\left\{\left(\mathbf{U}^{(q)},\mathbf{v}_1^{(q)},\ldots,\mathbf{v}_{N_{pan}}^{(q)}\right)\right\}$ generated by the algorithm are bounded. Furthermore, let $\left\{\left(\mathbf{F}',\boldsymbol{\varphi}_1',\ldots,\boldsymbol{\varphi}_{N_{pan}}'\right)\right\}$ be any accumulation point of set $\left\{\left(\mathbf{F}^{(q)},\boldsymbol{\varphi}_1^{(q)},\ldots,\boldsymbol{\varphi}_{N_{pan}}^{(q)}\right)\right\}$, then it must be a critical point of (4).

*Proof:* Proof is shown in Appendix C. ■

## III. Multi-hop RIS-assisted communication

### A. System Model

The second case that we address in this paper corresponds to a multi-hop scheme where, to help circumvent existing obstructions, the signal transmitted by the BS arrives at the receiver after being sequentially reflected by $N_{PAN}$ aiding RIS panels, as illustrated in Fig. 2. For this type of scenario, we still use the received signal model (1), but with the channel matrix defined as

$$\mathbf{H} = \mathbf{H}_{S,D} + \mathbf{H}_{N_{pan}+1} \prod_{i=1}^{N_{pan}} \mathbf{\Phi}_i \mathbf{H}_i \, . \tag{18}$$

In this expression $\mathbf{H}_1 \in \mathbb{C}^{N_{ris} \times N_{tx}}$ is the channel matrix between the transmitter and the first RIS panel, $\mathbf{H}_i \in \mathbb{C}^{N_{ris} \times N_{ris}}$ (i=2,…,$N_{pan}$) is the channel matrix between RIS panels $i$ and $i$+1 and $\mathbf{H}_{N_{pan}+1} \in \mathbb{C}^{N_{rx} \times N_{ris}}$ is the channel matrix between the $N_{pan}$th RIS panel and the receiver.

### *B. Precoder and RIS Optimization*

Since the received signal model in the multi-hop case is essentially the same as for the multiple parallel RIS scheme, we can also formulate the precoder and RIS design problem as (4a)-(4c) and solve it using Algorithm 1. However, due to the different channel matrix representation, the complex-valued gradient of $f\left(\mathbf{F}, \boldsymbol{\varphi}_1, ..., \boldsymbol{\varphi}_{N_{pan}}\right)$ will be different, as stated in the following lemma.

*Lemma 3:* Let $f\left(\mathbf{F}, \boldsymbol{\varphi}_1, ..., \boldsymbol{\varphi}_{N_{pan}}\right)$ be defined as in (4a) with the channel matrix represented as (18). Then, the corresponding complex-valued gradient with respect to $\mathbf{F}^*$ can be computed as (9) whereas the gradient with respect to $\boldsymbol{\varphi}_i^*$ ($i = 1, ..., N_{pan}$) can be computed as

$$\nabla_{\boldsymbol{\varphi}_i^*} f\left(\mathbf{F}, \boldsymbol{\varphi}_1, ..., \boldsymbol{\varphi}_{N_{pan}}\right) = -\frac{\rho}{P_n} \operatorname{diag}\left[ \bar{\mathbf{H}}_{i,D}{}^H \mathbf{H}\mathbf{F} \left( \mathbf{I}_{N_s} + \frac{\rho}{P_n} \mathbf{F}^H \mathbf{H}^H \mathbf{H}\mathbf{F} \right)^{-1} \mathbf{F}^H \bar{\mathbf{H}}_{S,i}{}^H \right] \tag{19}$$

with

$$\bar{\mathbf{H}}_{S,i} = \mathbf{H}_i \prod_{j=1}^{i-1} \mathbf{\Phi}_j \mathbf{H}_j \tag{20}$$

and

$$\bar{\mathbf{H}}_{i,D} = \mathbf{H}_{N_{pan}+1} \prod_{l=i+1}^{N_{pan}} \mathbf{\Phi}_l \mathbf{H}_l \, . \tag{21}$$

*Proof:* The derivation is straightforward by directly extending the approach presented in Appendix A for the multiple parallel RIS-assisted channel model.

Convergence of the JPR-mAPG algorithm can also be established for the multi-hop case through theorem 2, as long the gradients are still Lipschitz continuous. The following theorem states that this condition is still satisfied in this case.

*Theorem 3:* Function $f\left(\mathbf{F}, \boldsymbol{\varphi}_1, ..., \boldsymbol{\varphi}_{N_{pan}}\right)$ defined as in (4a), with the channel matrix represented as (18), has Lipschitz continuous gradients over feasible set (4b)-(4b), thus satisfying (11) with constant $L$ given by

$$L = \sqrt{\left(N_{pan} + 1\right) \max\left(b^2 + N_{pan} c^2, c^2 + N_{pan} d^2\right)}, \tag{22}$$

where

$$b = \frac{\rho}{P_n}\zeta^2\left(1+2\frac{N_s\rho}{P_n}\zeta^2\right) \quad (23)$$

$$c = 2\sqrt{N_s}\frac{\rho}{P_n}\zeta a^{N_{pan}-1}\left(\prod_{i=1}^{N_{pan}+1}\lambda_{\max}\left(\mathbf{H}_i\right)\right)\left(1+\frac{N_s\rho}{P_n}\zeta^2\right) \quad (24)$$

$$d = \frac{N_s\rho}{P_n}a^{2N_{pan}-2}\left(\prod_{i=1}^{N_{pan}+1}\lambda_{\max}^2\left(\mathbf{H}_i\right)\right)\left(1+2\frac{N_s\rho}{P_n}\zeta^2\right) \quad (25)$$

$$\zeta = \lambda_{\max}\left(\mathbf{H}_{S,D}\right)+a^{N_{pan}}\prod_{i=1}^{N_{pan}+1}\lambda_{\max}\left(\mathbf{H}_i\right). \quad (26)$$

*Proof:* Proof is similar to the one in Appendix B for Theorem 1. Due to the page limit the details are omitted here.

## IV. Numerical Results

In this section we evaluate the achievable rate of the proposed approach with the aid of Monte Carlo simulations. Two main scenarios are considered according to the two versions of the JPR-mAPG algorithm described previously namely, one based on multiple parallel RIS and one based on multi-hop.

### A. Channel Model

Even though the proposed approach is independent of a specific channel model, in this evaluation we adopt a clustered geometric channel [29], that is commonly assumed in mmWave [30] and sub-THz literature [31]. In the case of the channels between the transmitter, RIS and receiver, we consider that the RIS panels were properly located so that they consist of a LOS and a NLOS component, i.e., $\mathbf{H}_{S,i} = \mathbf{H}_{S,i}^{LOS} + \mathbf{H}_{S,i}^{NLOS}$ and $\mathbf{H}_{i,D} = \mathbf{H}_{i,D}^{LOS} + \mathbf{H}_{i,D}^{NLOS}$. We consider a spherical wave model for distances between transmitter, RIS and receiver smaller than the Fraunhofer distance defined as $D_F \triangleq 2L_{array}^2/\lambda$ [32], where $L_{array}$ is the maximum overall dimension of the array. Assuming unit normalized power radiation patterns along the directions of interest for both the antennas and RIS elements we can express the components of $\mathbf{H}_{S,i}^{LOS}$ as [33]

$$H_{S,i}^{LOS}\left(n,m\right) = \sqrt{\frac{G_{tx}A_{RIS}}{4\pi\left(d_{n,m}^{LOS,S,i}\right)^2}e^{-k_{abs}(f)d_{n,m}^{LOS,S,i}}}\, e^{-j2\pi d_{n,m}^{LOS,S,i}/\lambda}, \quad (27)$$

where $d_{n,m}^{LOS,S,i}$ denotes the distance of the LOS path between the $m^{\text{th}}$ transmit antenna element and the $n^{\text{th}}$ RIS element of the $i^{\text{th}}$ panel, $G_{tx}$ is the transmit antenna gain, $A_{RIS}$ is the area of a RIS element, and $k_{abs}(f)$ represents the molecular absorption coefficient at frequency *f*. Regarding the components of $\mathbf{H}_{S,i}^{NLOS}$, assuming $N_{ray}$ paths, they can be depicted as

$$H_{S,i}^{NLOS}(n,m)=\frac{1}{\sqrt{K_{Rice}}}\sum_{l=1}^{N_{ray}}\sqrt{\frac{G_{tx}A_{RIS}}{4\pi\left(d_{l,n,m}^{S,i}\right)^2}e^{-k_{abs}(f)d_{l,n,m}^{S,i}}}\left|\alpha_{l,n,m}^{S,i}\right|e^{-j2\pi d_{l,n,m}^{S,i}/\lambda}, \quad (28)$$

where $K_{Rice}$ defines the energy ratio between the LOS and NLOS components, $\alpha_{l,n,m}^{S,i}$ is the complex gain of the $l^{\text{th}}$ path between element *m* and *n* of the BS and $i^{\text{th}}$ RIS panel (normalized as $\sum_{l=1}^{N_{ray}}E\left[\left|\alpha_{l,n,m}^{RIS,S}\right|^2\right]=1$) and $d_{l,n,m}^{S,i}$ denotes the respective distance. The components of $\mathbf{H}_{i,D}^{LOS}$ and $\mathbf{H}_{i,D}^{NLOS}$ can be written using expressions similar to (*27*) and (28).

When the distances between transmitter, RIS and receiver are larger than the Fraunhofer distance, we adopt a planar wave model. In this case, the NLOS channel matrix is approximated as

$$\mathbf{H}_{S,i}^{NLOS}=\sqrt{\frac{\beta_{NLOS}^{S,i}}{K_{Rice}}}\sum_{l=1}^{N_{ray}}\alpha_l^{S,i}\mathbf{a}_{RIS}\left(\phi_l^{i\leftarrow S},\theta_l^{i\leftarrow S}\right)\mathbf{a}_S\left(\phi_l^{S\rightarrow i},\theta_l^{S\rightarrow i}\right)^H, \quad (29)$$

where $\mathbf{a}_{RIS}\left(\phi_l^{i\leftarrow S},\theta_l^{i\leftarrow S}\right)$ and $\mathbf{a}_S\left(\phi_l^{S\rightarrow i},\theta_l^{S\rightarrow i}\right)$ denote the array responses of the BS and RIS panel at the respective angles of departure (AoD) and angles of arrival (AoA) $\left(\phi_l^{i\leftarrow S},\theta_l^{i\leftarrow S}\right)$ and $\left(\phi_l^{S\rightarrow i},\theta_l^{S\rightarrow i}\right)$ ($\phi$ denotes the azimuth and $\theta$ the elevation). $\beta_{NLOS}^{S,i}$ represents the path loss, which we express as

$$\beta_{NLOS}^{S,i}=\frac{G_{tx}A_{RIS}}{4\pi\left(d_{S\leftrightarrow RIS}^{i}\right)^{\gamma}}e^{-k_{abs}(f)d_{S\leftrightarrow RIS}^{i}}, \quad (30)$$

where $d_{S\leftrightarrow RIS}^{i}$ denotes the distance between the center of the transmitter array and the center of the $i^{\text{th}}$ RIS panel and $\gamma$ is the path loss exponent. A similar expression is used for the LOS matrix (in this case with only one path). For the direct channel between the BS and the user, we consider that due to the presence of many obstructing obstacles it will typically consist solely of NLOS components and, therefore, can be expressed similarly to (28) or (29).

In all simulations uniform planar arrays (UPAs) with $\sqrt{N_{tx}}\times\sqrt{N_{tx}}$, $\sqrt{N_{rx}}\times\sqrt{N_{rx}}$ and $\sqrt{N_{ris}}\times\sqrt{N_{ris}}$ elements are adopted at the transmitter, receiver and RIS, respectively. In this case, the steering vectors for the transmitter array are given by

$$\mathbf{a}_S\left(\phi_l^{S\rightarrow i},\theta_l^{S\rightarrow i}\right)=\left[1,\ldots,e^{j\frac{2\pi}{\lambda}d_S\left(p\sin\phi_l^{S\rightarrow i}\sin\theta_l^{S\rightarrow i}+q\cos\theta_l^{S\rightarrow i}\right)},\ldots,e^{j\frac{2\pi}{\lambda}d_S\left(\left(\sqrt{N_{tx}}-1\right)\sin\phi_l^{S\rightarrow i}\sin\theta_l^{S\rightarrow i}+\left(\sqrt{N_{tx}}-1\right)\cos\theta_l^{S\rightarrow i}\right)}\right], \quad (31)$$

where *p*,*q* are the antenna indices, and $d_S$ is the inter-element spacing. A similar expression is adopted for the RIS and receiver arrays. In the simulations, the inter-element spacing of the transmitter and receiver arrays, as well as the distance between the centers of adjacent elements of the RIS are always set as $\lambda/2$ ,whereas the sizes of the RIS elements are $\lambda/2\times\lambda/2$ .

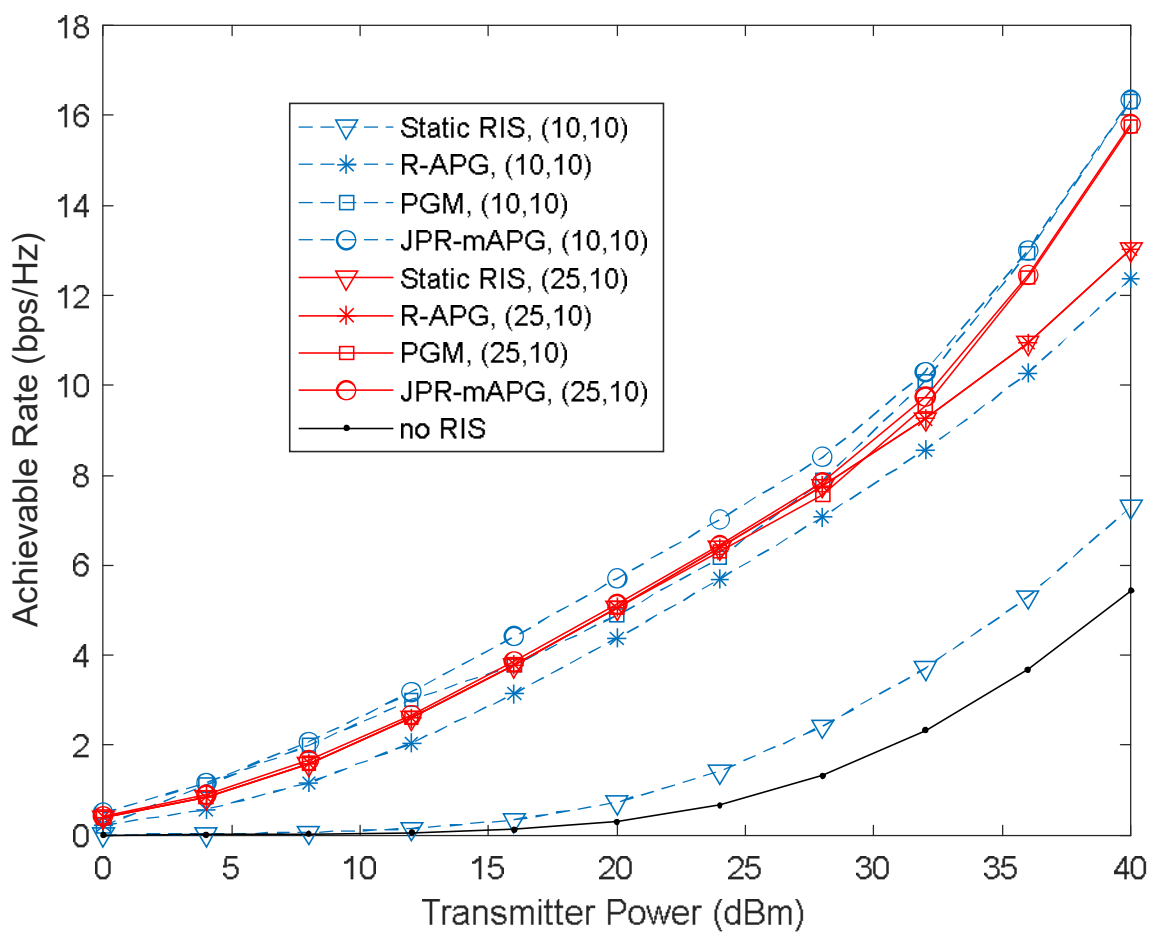

Fig. 3. Achievable rate versus transmitted power with different algorithms and different locations of a single RIS panel. Scenario with: $N_s$=3, $N_{tx}$=64, $N_{rx}$=16 and $N_{ris}$=256.

## *B. Multiple parallel RIS Scenario*

In the first scenario we consider a large indoor environment (office/shopping mall) where one or more RISs are deployed in the vicinity of the BS. The operating frequency is set to 28 GHz with an occupied bandwidth of 800 MHz. We assume that the amplitude of the reflection coefficients is $a$ =1. Based on [34] we apply a path loss exponent of 1.90 for the LOS propagation and 4.39 for the NLOS propagation. The number of propagation paths is set to $N_{ray}$=10, with the azimuth and elevation AoD/AoA uniformly random distributed in $(-\pi,\pi)$ and $(-\pi/2,\pi/2)$, respectively [35]. For the indirect link through the RIS we apply $K_{Rice}$=10, whereas the direct link has no LOS component ($K_{Rice}$=0). All simulated results are averaged over 500 random channel realizations. To simplify the setups, we work in a two-dimensional coordinate system (the third coordinate is assumed to be the same for the BS, RIS panels and receiver) with the BS located at (0,0) m.

Fig. 3 shows the achievable rate as a function of the transmitted power with $N_s$=3, $N_{tx}$=64, $N_{rx}$=16, $N_{ris}$=256 and only one panel. The evaluation is performed with the RIS panel in two different positions: (10,10) m and (25,10) m. As benchmarks, we include curves for a static RIS (acting as a simple mirror), the RIS-only APG algorithm (R-APG) from [12] and the PGM approach from [17]. As a reference we also include the case of no RIS. It can be seen that the adoption of a RIS can substantially improve the achievable rate over a transmission without RIS, with the proposed JPR-mAPG algorithm providing a 5-fold improvement at $P_{tx}$=30 dBm. We can also observe that this gain is higher when the RIS is closer to the BS (or closer to the user, even though these curves were not included). An exception occurs for the case of static RIS since it is acting as a simple mirror and at (25,10) m it is located at a position and with an orientation where the incidence angle of the LOS ray between the BS and RIS coincides with the reflected angle of the LOS ray between the RIS and the receiver. Comparing the different RIS approaches, we can see that both JPR-mAPG and PGM achieve similar results at high transmission powers with the former one obtaining better rates at moderate power levels. R-APG displays a somewhat inferior performance as it only tries to optimize the RIS matrix after the precoder is computed.

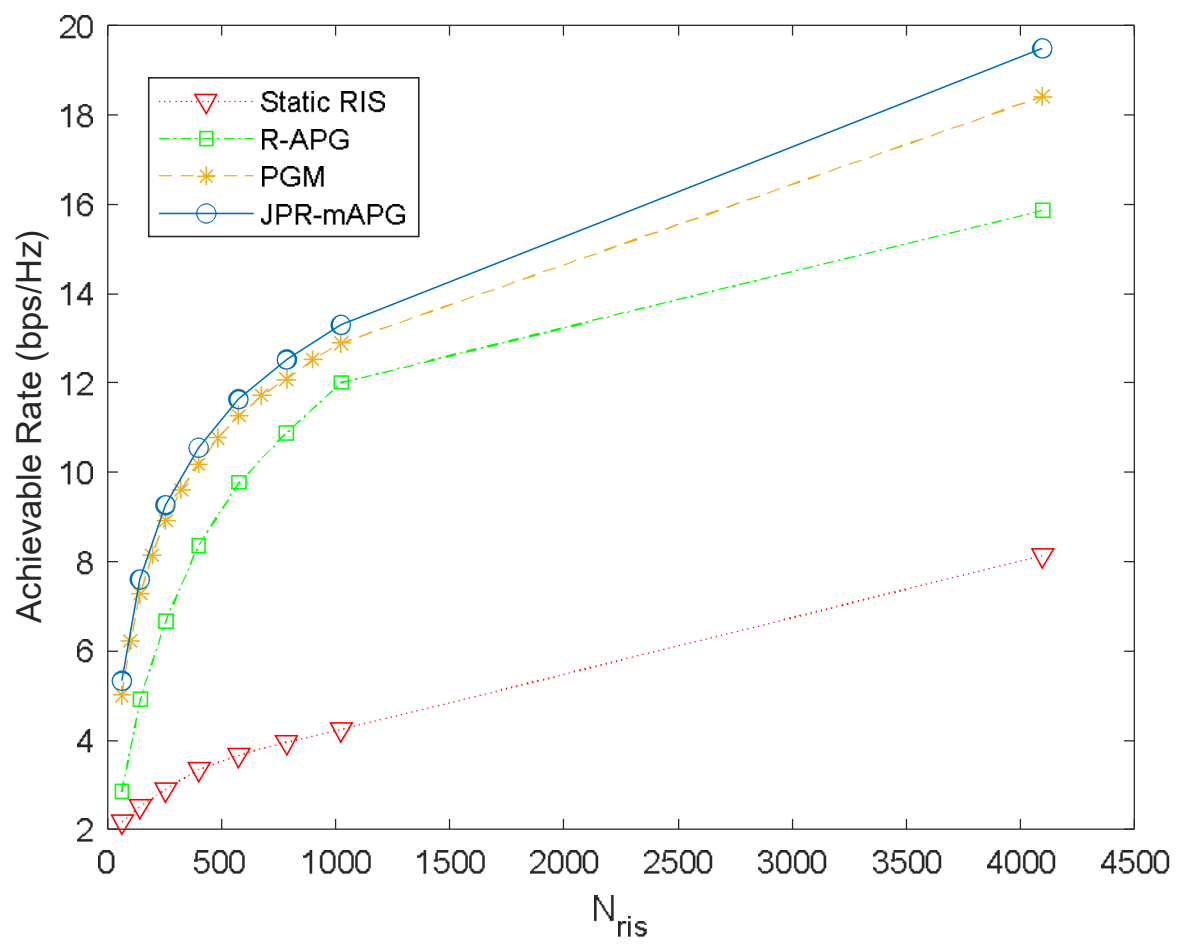

Fig. 4. Achievable rate versus $N_{ris}$ with different algorithms and a single RIS panel. Scenario with: $N_s$=3, $N_{tx}$=64, $N_{rx}$=16 and $P_{tx}$=30 dBm.

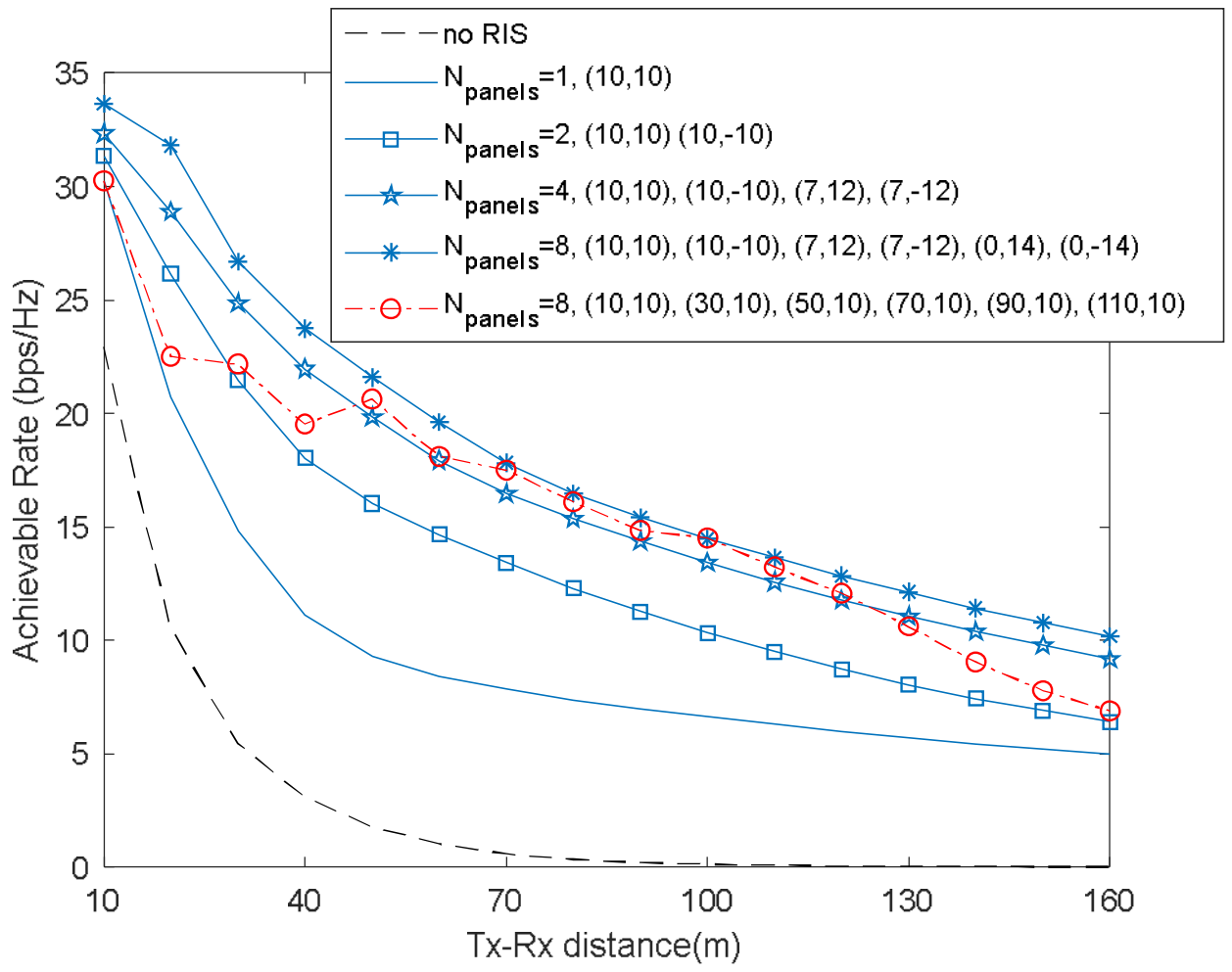

Fig. 5. Achievable rate versus distance of the proposed JPR-mAPG algorithm with different multi-RIS assisted configurations. Scenario with $N_s$=3, $N_{tx}$=64, $N_{rx}$=16, $N_{ris}$=256.

Keeping basically the same setup but with $P_{tx}$=30 dBm and the RIS positioned at (10,10) m, Fig. 4 evaluates the impact of the number of RIS elements using different algorithms. We can observe that the achievable rate clearly improves with the number of RIS elements, with the curves showing a higher slope for low $N_{ris}$ values, which gradually decreases as the number of elements grows. Similarly to the previous figure, all the evaluated RIS optimization algorithms provide substantial gains over a static RIS, with the proposed JPR-mAPG approach achieving the best results.

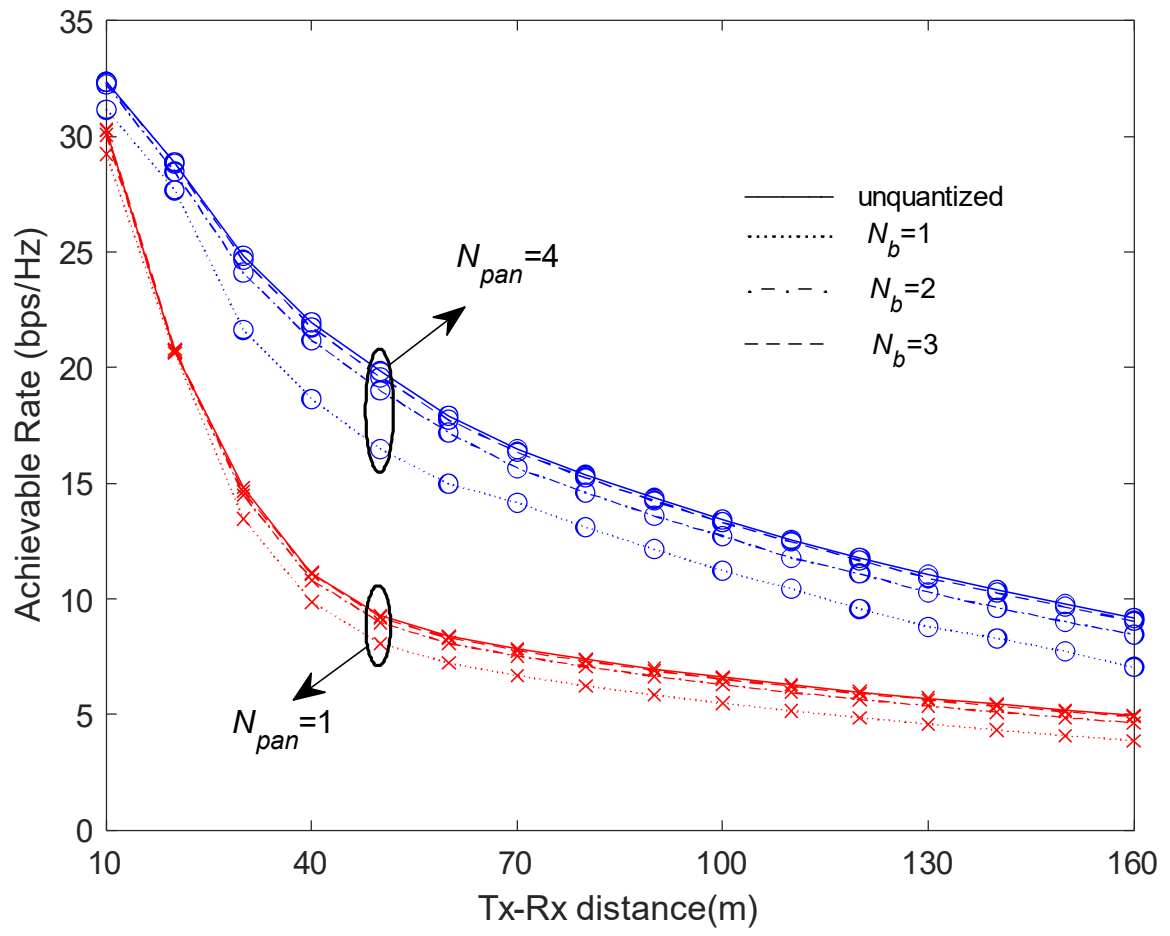

Fig. 6. Achievable rate versus distance of the proposed JPR-mAPG algorithm considering different multi-RIS assisted configurations with discrete phase shifts. Scenario with $N_s$=3, $N_{tx}$=64, $N_{rx}$=16, $N_{ris}$=256.

Considering the same scenario, Fig. 5 shows the achievable rate versus transmitter-receiver distance, depicting the effect of using multiple RIS panels combined with the proposed algorithm. As expected, the adoption of multiple RIS panels can improve the rates even further. However, similarly to what was observed in the single panel case, the location of the various panels can have a significant impact on the results. Considering the case $N_{pan}$=8, it can be seen that spreading the panels along the path between the transmitter and receiver instead of locating them all in the vicinity of the BS can result in a reduced number of panels being effective in aiding the communication depending on the position of the receiver.

In practical implementations of RIS-assisted systems, only a finite number of phase shifts can be supported in the RIS elements [36]. Therefore, in Fig. 6 we evaluate the impact of using discrete phase shifts at the RIS, considering the same setup of the previous figures. The proposed JPR-mAPG algorithm is applied with a final projection of each individual unquantized phase onto the set of discrete phase shifts. In the legend, $N_b$ denotes the number of quantization bits. We can observe that the performance becomes only slightly more sensitive to the phase quantization effect when using multiple panels. For example, there is a degradation around 17% at a distance of 50 m with 4 panels when using one bit quantization whereas with a single panel the degradation is 13%.

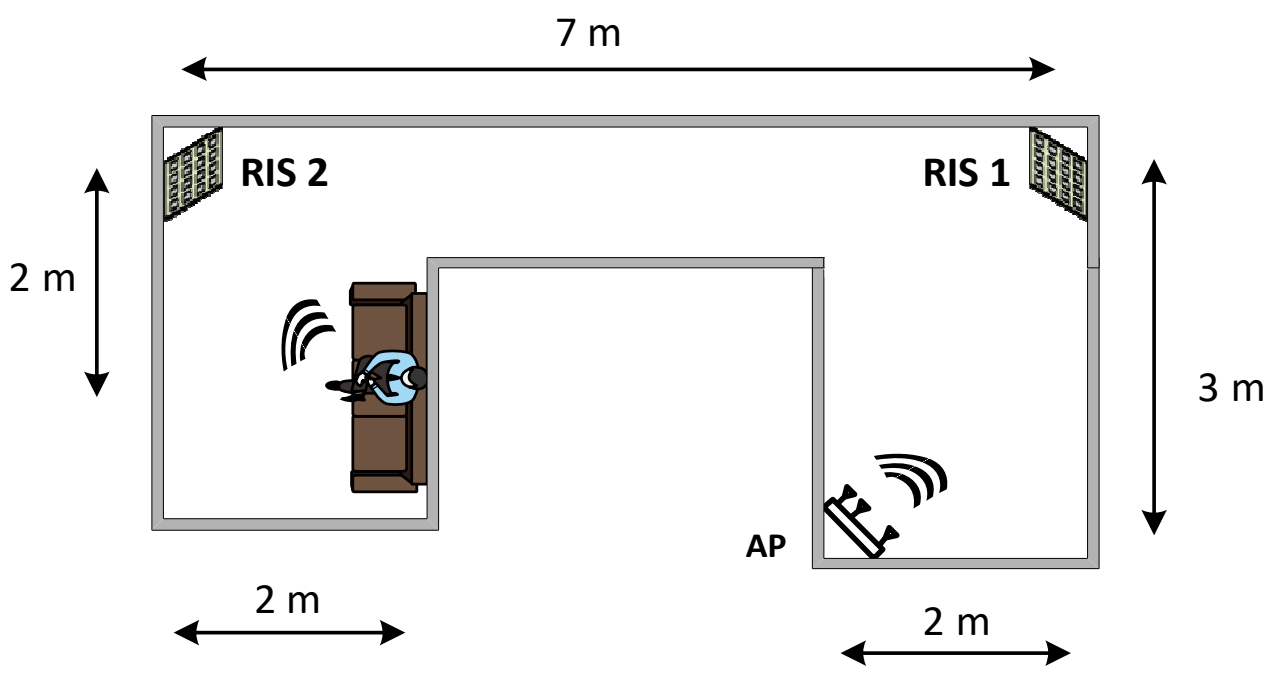

Fig. 7. Overhead layout of the simulated multi-hop scenario.

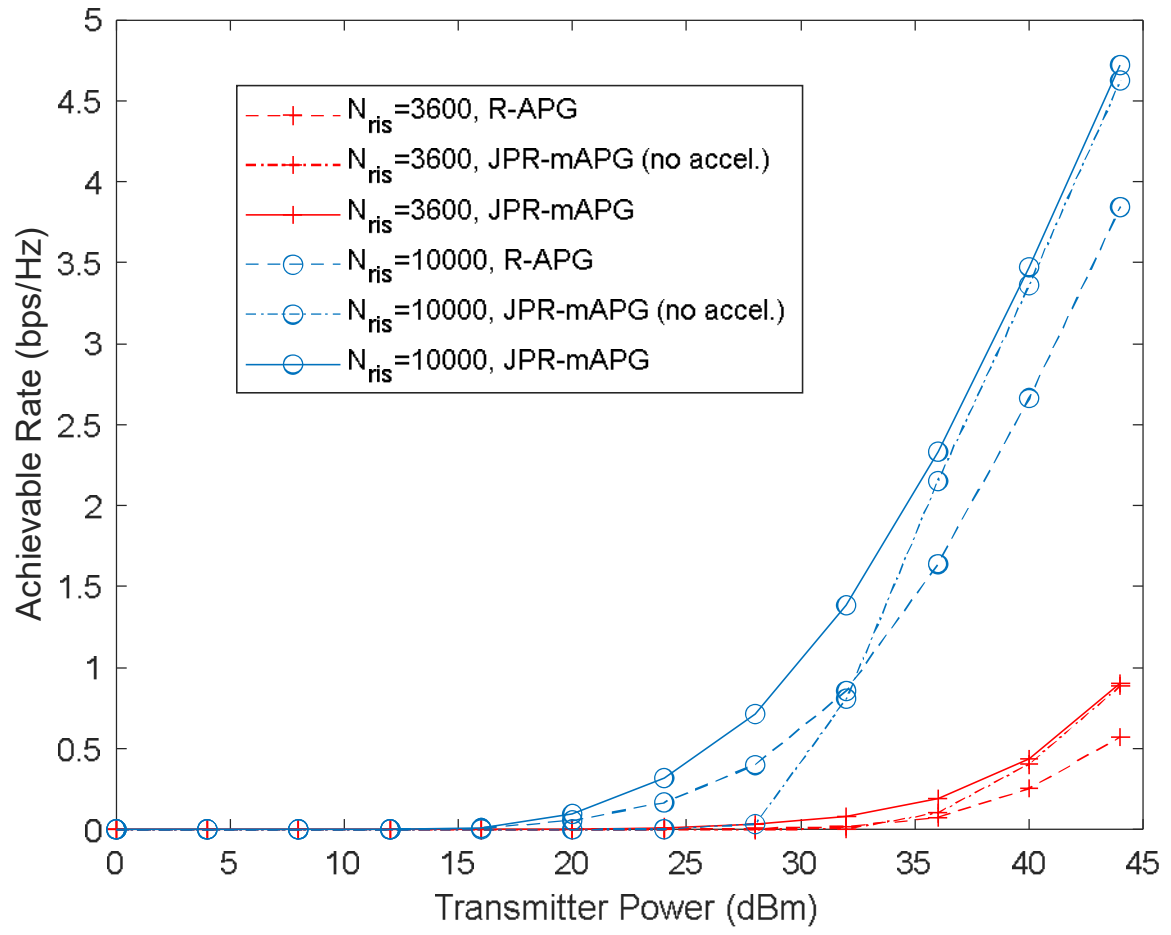

Fig. 8. Achievable rate versus distance in a two-hop RIS-assisted setting. Scenario with $N_s$=1, $N_{tx}$=1024, $N_{rx}$=36.

Still, the difference to the unquantized results reduces to a minor value with only 2 bits, and with 3 bits the performance basically matches the ideal unquantized curves.

### *C. Multi-hop RIS-assisted Scenario*

In the second scenario we consider that the transmission between BS and user is supported with the aid of two RIS panels, establishing a two-hop communication scheme. The layout with the positions of BS, user and RIS panels is presented in Fig. 7. In this case the system is operating at the sub-THz band with the frequency set to 100 GHz and the occupied bandwidth kept at 800 MHz. Due to the presence of walls/obstacles, we assume that there is no direct link between BS and user. For all the links through the RIS panels we consider LOS propagation with a path loss exponent of 2.05 [34], with the number of propagation paths set to $N_{ray}$=3 and $K_{Rice}$=10. The other parameters are configured as $N_s$=1, $N_{tx}$=1024, $N_{rx}$=36. Fig. 8 shows the achievable rate versus transmitted power of the proposed JPR-mAPG approach as well as of the R-APG algorithm. A version of the JPR-mAPG without acceleration is also included which corresponds to applying only steps 6 and 10 in Algorithm 1. We can see that working in a higher frequency band and with a multi-hop scheme requires the adoption of RIS panels with a higher number of elements (even though not necessarily larger due to the smaller wavelength) to achieve acceptable rates. Similarly to what was observed in the previous scenario, adopting a joint design of the precoder and RIS, as achieves better performances than applying an algorithm that only optimizes the RIS panels. Furthermore, there is also a clear improvement with the adoption of the extrapolation step in the JPR-mAPG instead of a direct projected gradient step (JPR-mAPG without acceleration).

## V. Conclusions

In this paper we addressed the joint design of the transmit precoder and RIS phases within the context of multi-stream MIMO communications assisted by multiple RIS panels. Aiming at solving the nonconvex problem of maximizing the achievable rate, we derived an algorithm based on the mAPG method that can be tailored for the case of parallel RISs as well as for multi-hop RIS-assisted scenarios. The mAPG approach relies on an extrapolation step for improving the convergence speed and a set of monitoring variables that ensure the sufficient descent of the algorithm. Detailed convergence analysis was presented which included expressions for the step size that guarantee convergence to a critical point. Simulation results showed that the proposed JPR-mAPG algorithm can often achieve higher rates than other benchmarked approaches. Furthermore, it was observed that the use of properly located multiple RIS panels operating in parallel can have a substantial impact on the achieved rates, whereas the adoption of several RIS panels in a multi-hop configuration can be an effective approach for providing coverage in the case of severe blockage effects.

## Appendix A

### Proof of Lemma 1

The following derivations for the complex-valued gradient of $f\left(\mathbf{F},\boldsymbol{\varphi}_1,...,\boldsymbol{\varphi}_{N_{pan}}\right)$ with respect to $\mathbf{F}^*$ and $\boldsymbol{\varphi}_i^*$ $(i=1,...,N_{pan})$ are based on the procedure described in [37]. First, we derive the complex differential of $f\left(\mathbf{F},\boldsymbol{\varphi}_1,...,\boldsymbol{\varphi}_{N_{pan}}\right)$ with respect to $\mathbf{F}^*$ as

$$df = -\text{Tr}\left\{\frac{\rho}{P_n}\left(\mathbf{I}_{N_s}+\frac{\rho}{P_n}\mathbf{F}^H\mathbf{H}^H\mathbf{H}\mathbf{F}\right)^{-1}\left(\mathbf{F}^H\mathbf{H}^H\mathbf{H}d\mathbf{F}+d\mathbf{F}^H\mathbf{H}^H\mathbf{H}\mathbf{F}\right)\right\}$$

$$= -\mathrm{Tr}\left\{ \frac{\rho}{P_n}\left(\mathbf{I}_{N_s} + \frac{\rho}{P_n}\mathbf{F}^H\mathbf{H}^H\mathbf{H}\mathbf{F}\right)^{-1}\mathbf{F}^H\mathbf{H}^H\mathbf{H}d\mathbf{F} + \frac{\rho}{P_n}\left(\mathbf{H}^H\mathbf{H}\mathbf{F}\left(\mathbf{I}_{N_s} + \frac{\rho}{P_n}\mathbf{F}^H\mathbf{H}^H\mathbf{H}\mathbf{F}\right)^{-1}\right)^T d\mathbf{F}^* \right\} \quad (32)$$

where we used the relation (Table 3.1 in [37])

$$d\left(\ln\det\left(\mathbf{X}\right)\right) = \mathrm{Tr}\left\{\mathbf{X}^{-1}d\mathbf{X}\right\} \quad . \quad (33)$$

Then, using Table 3.2 from [37] we obtain (9). Regarding the complex differential of $f\left(\mathbf{F}, \boldsymbol{\varphi}_1, \ldots, \boldsymbol{\varphi}_{N_{pan}}\right)$ with respect to $\boldsymbol{\varphi}_i^*$, it can be written as

$$df = -\mathrm{Tr}\left\{ \frac{\rho}{P_n}\left(\mathbf{I}_{N_s} + \frac{\rho}{P_n}\mathbf{F}^H\mathbf{H}^H\mathbf{H}\mathbf{F}\right)^{-1}\left(\mathbf{F}^H\mathbf{H}^H d\mathbf{H}\mathbf{F} + \mathbf{F}^H d\mathbf{H}^H\mathbf{H}\mathbf{F}\right) \right\}. \quad (34)$$

Noting that

$$d\mathbf{H} = \mathbf{H}_{i,D}\mathrm{diag}\left(d\boldsymbol{\varphi}_i\right)\mathbf{H}_{S,i} \quad (35)$$

and

$$\mathrm{Tr}\left\{\mathbf{Z}\mathrm{diag}\left(d\boldsymbol{\varphi}_i\right)\right\} = \mathrm{diag}\left(\mathbf{Z}\right)^T d\boldsymbol{\varphi}_i \quad (36)$$

then we obtain

$$\begin{aligned} df = &-\frac{\rho}{P_n}\mathrm{diag}\left[\mathbf{H}_{S,i}\mathbf{F}\left(\mathbf{I}_{N_s} + \frac{\rho}{P_n}\mathbf{F}^H\mathbf{H}^H\mathbf{H}\mathbf{F}\right)^{-1}\mathbf{F}^H\mathbf{H}^H\mathbf{H}_{i,D}\right]^T d\boldsymbol{\varphi}_i \\ &-\frac{\rho}{P_n}\mathrm{diag}\left[\mathbf{H}_{i,D}{}^H\mathbf{H}\mathbf{F}\left(\mathbf{I}_{N_s} + \frac{\rho}{P_n}\mathbf{F}^H\mathbf{H}^H\mathbf{H}\mathbf{F}\right)^{-1}\mathbf{F}^H\mathbf{H}_{S,i}{}^H\right]^T d\boldsymbol{\varphi}_i^* . \end{aligned} \quad (37)$$

Again, using Table 3.2 from [37] gives the gradient expression in (10). ■

## APPENDIX B

## PROOF OF THEOREM 1

The following inequalities will be useful for the proof. For matrices **A** and **B**, we have

$$\left\|\mathbf{A}\mathbf{B}\right\| \le \left\|\mathbf{A}\right\|\left\|\mathbf{B}\right\| \quad (38)$$

for the Frobenius and the 2-norm. Also

$$\left\|\mathbf{A}\right\|_2 \le \left\|\mathbf{A}\right\|_F \quad (39)$$

and

$$\left\|\mathbf{A}\mathbf{B}\right\|_F \le \left\|\mathbf{A}\right\|_2\left\|\mathbf{B}\right\|_F . \quad (40)$$

Before proving Theorem 1, we first introduce the following Lemma which will be applied in several steps of the proof.

*Lemma 2:* Let $\left\{\mathbf{\Psi}_k^{(i)}\right\}$ with $i$=1,2, and $k$=1,..,$N$ ($N$≥2) be a set of matrices. Then we have the following inequality

$$\left\|\prod_{k=1}^{N}\mathbf{\Psi}_k^{(2)}-\prod_{k=1}^{N}\mathbf{\Psi}_k^{(1)}\right\|_F\leq\sum_{l=1}^{N}\left(\prod_{m=l+1}^{N}\left\|\mathbf{\Psi}_m^{(1)}\right\|_2\right)\left\|\mathbf{\Psi}_l^{(2)}-\mathbf{\Psi}_l^{(1)}\right\|_F\left(\prod_{k=1}^{l-1}\left\|\mathbf{\Psi}_k^{(2)}\right\|_2\right)\tag{41}$$

*Proof:* We prove by induction on $N$. For $N$=2 we can write:

$$\begin{aligned}\left\|\mathbf{\Psi}_2^{(2)}\mathbf{\Psi}_1^{(2)}-\mathbf{\Psi}_2^{(1)}\mathbf{\Psi}_1^{(1)}\right\|_F&\leq\left\|\mathbf{\Psi}_2^{(2)}-\mathbf{\Psi}_2^{(1)}\right\|_F\left\|\mathbf{\Psi}_1^{(2)}\right\|_2+\left\|\mathbf{\Psi}_2^{(1)}\right\|_2\left\|\mathbf{\Psi}_1^{(2)}-\mathbf{\Psi}_1^{(1)}\right\|_F\\&\leq\sum_{l=1}^{2}\left(\prod_{m=l+1}^{2}\left\|\mathbf{\Psi}_m^{(1)}\right\|_2\right)\left\|\mathbf{\Psi}_l^{(2)}-\mathbf{\Psi}_l^{(1)}\right\|_F\left(\prod_{k=1}^{l-1}\left\|\mathbf{\Psi}_k^{(2)}\right\|_2\right)\end{aligned}\tag{42}$$

thus showing that the inequality is valid. For $N$+1 we can write

$$\begin{aligned}\left\|\prod_{k=1}^{N+1}\mathbf{\Psi}_k^{(2)}-\prod_{k=1}^{N+1}\mathbf{\Psi}_k^{(1)}\right\|_F&=\left\|\mathbf{\Psi}_{N+1}^{(2)}\prod_{k=1}^{N}\mathbf{\Psi}_k^{(2)}-\mathbf{\Psi}_{N+1}^{(1)}\prod_{k=1}^{N}\mathbf{\Psi}_k^{(1)}\right\|_F\\&\leq\left\|\mathbf{\Psi}_{N+1}^{(2)}-\mathbf{\Psi}_{N+1}^{(1)}\right\|_F\prod_{k=1}^{N}\left\|\mathbf{\Psi}_k^{(2)}\right\|_2+\left\|\mathbf{\Psi}_{N+1}^{(1)}\right\|_2\left\|\prod_{k=1}^{N}\mathbf{\Psi}_k^{(2)}-\prod_{k=1}^{N}\mathbf{\Psi}_k^{(1)}\right\|_F.\end{aligned}\tag{43}$$

Assume that inequality (41) holds for $N$. Then, we can rewrite (43) as

$$\begin{aligned}&\left\|\prod_{k=1}^{N+1}\mathbf{\Psi}_k^{(2)}-\prod_{k=1}^{N+1}\mathbf{\Psi}_k^{(1)}\right\|_F\\&\leq\left\|\mathbf{\Psi}_{N+1}^{(2)}-\mathbf{\Psi}_{N+1}^{(1)}\right\|_F\prod_{k=1}^{N}\left\|\mathbf{\Psi}_k^{(2)}\right\|_2+\left\|\mathbf{\Psi}_{N+1}^{(1)}\right\|_2\sum_{l=1}^{N}\left(\prod_{m=l+1}^{N}\left\|\mathbf{\Psi}_m^{(1)}\right\|_2\right)\left\|\mathbf{\Psi}_l^{(2)}-\mathbf{\Psi}_l^{(1)}\right\|_F\left(\prod_{k=1}^{l-1}\left\|\mathbf{\Psi}_k^{(2)}\right\|_2\right).\end{aligned}\tag{44}$$

Noting that

$$\left\|\mathbf{\Psi}_{N+1}^{(2)}-\mathbf{\Psi}_{N+1}^{(1)}\right\|_F\prod_{k=1}^{N}\left\|\mathbf{\Psi}_k^{(2)}\right\|_2=\left(\prod_{m=N+1+1}^{N+1}\left\|\mathbf{\Psi}_m^{(1)}\right\|_2\right)\left\|\mathbf{\Psi}_{N+1}^{(2)}-\mathbf{\Psi}_{N+1}^{(1)}\right\|_F\prod_{k=1}^{N+1-1}\left\|\mathbf{\Psi}_k^{(2)}\right\|_2$$

we can finally write

$$\left\|\prod_{k=1}^{N+1}\mathbf{\Psi}_k^{(2)}-\prod_{k=1}^{N+1}\mathbf{\Psi}_k^{(1)}\right\|_F\leq\sum_{l=1}^{N+1}\left(\prod_{m=l+1}^{N+1}\left\|\mathbf{\Psi}_m^{(1)}\right\|_2\right)\left\|\mathbf{\Psi}_l^{(2)}-\mathbf{\Psi}_l^{(1)}\right\|_F\left(\prod_{k=1}^{l-1}\left\|\mathbf{\Psi}_k^{(2)}\right\|_2\right)\tag{45}$$

which shows that the inequality is also valid. ■

We note that inequality (41) can also be written using the spectral norm since $\left\|\mathbf{X}\right\|_2\leq\left\|\mathbf{X}\right\|_F$.

*Proof of Theorem 1:* First we define the following concatenated matrices and vectors

$$\tilde{\boldsymbol{\varphi}}=\left[\boldsymbol{\varphi}_1{}^T,\ldots,\boldsymbol{\varphi}_{N_{pan}}{}^T\right]^T,\tag{46}$$

$$\tilde{\mathbf{H}}_{R,D}=\left[\mathbf{H}_{1,D}\ldots\mathbf{H}_{N_{pan},D}\right],\tag{47}$$

$$\tilde{\mathbf{H}}_{S,R}=\left[\mathbf{H}_{S,1}{}^T\ldots\mathbf{H}_{S,N_{pan}}{}^T\right]^T\tag{48}$$

and $\tilde{\mathbf{\Phi}}=\mathrm{diag}\left(\tilde{\boldsymbol{\varphi}}\right)$ which enable us to rewrite (2) in a more compact form

$$\mathbf{H}=\mathbf{H}_{S,D}+\tilde{\mathbf{H}}_{R,D}\tilde{\mathbf{\Phi}}\tilde{\mathbf{H}}_{S,R}.\tag{49}$$

This also allows us to work with the concatenated gradient vector

$$\nabla_{\tilde{\boldsymbol{\varphi}}^*} f\left(\mathbf{F},\tilde{\boldsymbol{\varphi}}\right)=\left[\nabla_{\boldsymbol{\varphi}_1^*} f\left(\mathbf{F},\boldsymbol{\varphi}_1,\ldots,\boldsymbol{\varphi}_{N_{pan}}\right)^T,\ldots,\nabla_{\boldsymbol{\varphi}_{N_{pan}}^*} f\left(\mathbf{F},\boldsymbol{\varphi}_1,\ldots,\boldsymbol{\varphi}_{N_{pan}}\right)^T\right]^T \tag{50}$$

and write

$$\begin{aligned}&\left\|\nabla_{\tilde{\boldsymbol{\varphi}}^*} f\left(\mathbf{F}^{(q_2)},\tilde{\boldsymbol{\varphi}}^{(q_2)}\right)-\nabla_{\tilde{\boldsymbol{\varphi}}^*} f\left(\mathbf{F}^{(q_1)},\tilde{\boldsymbol{\varphi}}^{(q_1)}\right)\right\|_2\\ &=\frac{\rho}{P_n}\left\|\operatorname{diag}\left(\underset{i=1,\ldots,N_{pan}}{\operatorname{blkdiag}}\left\{\left(\mathbf{H}_{i,D}\right)^H\mathbf{H}^{(q_2)}\mathbf{F}^{(q_2)}\mathbf{K}\left(\mathbf{F}^{(q_2)},\tilde{\boldsymbol{\varphi}}^{(q_2)}\right)\mathbf{F}^{(q_2)H}\left(\mathbf{H}_{S,i}\right)^H\right.\right.\right.\\ &\left.\left.\left.-\left(\mathbf{H}_{i,D}\right)^H\mathbf{H}^{(q_1)}\mathbf{F}^{(q_1)}\mathbf{K}\left(\mathbf{F}^{(q_1)},\tilde{\boldsymbol{\varphi}}^{(q_1)}\right)\mathbf{F}^{(q_1)H}\left(\mathbf{H}_{S,i}\right)^H\right\}\right)\right\|_2.\end{aligned} \tag{51}$$

with $\mathbf{K}\left(\mathbf{F},\tilde{\boldsymbol{\varphi}}\right)=\left(\mathbf{I}_{N_s}+\frac{\rho}{P_n}\mathbf{F}^H\mathbf{H}^H\mathbf{H}\mathbf{F}\right)^{-1}$. We can rewrite this expression as

$$\begin{aligned}\left\|\nabla_{\tilde{\boldsymbol{\varphi}}^*} f\left(\mathbf{F}^{(q_2)},\tilde{\boldsymbol{\varphi}}^{(q_2)}\right)-\nabla_{\tilde{\boldsymbol{\varphi}}^*} f\left(\mathbf{F}^{(q_1)},\tilde{\boldsymbol{\varphi}}^{(q_1)}\right)\right\|_2 &\le\frac{\rho}{P_n}\sum_{i=1}^{N_{pan}}\left\|\left(\mathbf{H}_{i,D}\right)^H\mathbf{H}^{(q_2)}\mathbf{F}^{(q_2)}\mathbf{K}\left(\mathbf{F}^{(q_2)},\tilde{\boldsymbol{\varphi}}^{(q_2)}\right)\mathbf{F}^{(q_2)H}\left(\mathbf{H}_{S,i}\right)^H\right.\\ &\left.-\left(\mathbf{H}_{i,D}\right)^H\mathbf{H}^{(q_1)}\mathbf{F}^{(q_1)}\mathbf{K}\left(\mathbf{F}^{(q_1)},\tilde{\boldsymbol{\varphi}}^{(q_1)}\right)\mathbf{F}^{(q_1)H}\left(\mathbf{H}_{S,i}\right)^H\right\|_F.\end{aligned} \tag{52}$$

It can be easily shown that $\left\|\mathbf{K}\left(\mathbf{F},\tilde{\boldsymbol{\varphi}}\right)\right\|_2\le 1$ and $\left\|\boldsymbol{\Phi}_i\right\|_2=a$. Using Lemma 2 we can write

$$\left\|\mathbf{H}^{(q_2)}-\mathbf{H}^{(q_1)}\right\|_F\le\lambda_{\max}\left(\tilde{\mathbf{H}}_{R,D}\right)\lambda_{\max}\left(\tilde{\mathbf{H}}_{S,R}\right)\left\|\tilde{\boldsymbol{\varphi}}^{(q_2)}-\tilde{\boldsymbol{\varphi}}^{(q_1)}\right\|_2, \tag{53}$$

$$\begin{aligned}\left\|\mathbf{K}\left(\mathbf{F}^{(q_2)},\tilde{\boldsymbol{\varphi}}^{(q_2)}\right)-\mathbf{K}\left(\mathbf{F}^{(q_1)},\tilde{\boldsymbol{\varphi}}^{(q_1)}\right)\right\|_F&\le\left\|\mathbf{K}\left(\mathbf{F}^{(q_2)},\tilde{\boldsymbol{\varphi}}^{(q_2)}\right)^{-1}-\mathbf{K}\left(\mathbf{F}^{(q_1)},\tilde{\boldsymbol{\varphi}}^{(q_1)}\right)^{-1}\right\|_F\\ &\le\frac{2\rho}{P_n}\left(\sqrt{N_s}\zeta^2\left\|\mathbf{F}^{(q_2)}-\mathbf{F}^{(q_1)}\right\|_F+N_s\zeta\lambda_{\max}\left(\tilde{\mathbf{H}}_{R,D}\right)\lambda_{\max}\left(\tilde{\mathbf{H}}_{S,R}\right)\left\|\tilde{\boldsymbol{\varphi}}^{(q_2)}-\tilde{\boldsymbol{\varphi}}^{(q_1)}\right\|_2\right).\end{aligned} \tag{54}$$

Applying these inequalities and Lemma 2 to (*52*) results

$$\left\|\nabla_{\tilde{\boldsymbol{\varphi}}^*} f\left(\mathbf{F}^{(q_2)},\tilde{\boldsymbol{\varphi}}^{(q_2)}\right)-\nabla_{\tilde{\boldsymbol{\varphi}}^*} f\left(\mathbf{F}^{(q_1)},\tilde{\boldsymbol{\varphi}}^{(q_1)}\right)\right\|_2\le d\left\|\mathbf{F}^{(q_2)}-\mathbf{F}^{(q_1)}\right\|_F+e\left\|\tilde{\boldsymbol{\varphi}}^{(q_2)}-\tilde{\boldsymbol{\varphi}}^{(q_1)}\right\|_2. \tag{55}$$

Similarly, we can also use Lemma 2 and write the following inequality

$$\left\|\nabla_{\mathbf{F}^*} f\left(\mathbf{F}^{(q_2)},\tilde{\boldsymbol{\varphi}}^{(q_2)}\right)-\nabla_{\mathbf{F}^*} f\left(\mathbf{F}^{(q_1)},\tilde{\boldsymbol{\varphi}}^{(q_1)}\right)\right\|_F=b\left\|\mathbf{F}^{(q_2)}-\mathbf{F}^{(q_1)}\right\|_F+c\left\|\tilde{\boldsymbol{\varphi}}^{(q_2)}-\tilde{\boldsymbol{\varphi}}^{(q_1)}\right\|_2. \tag{56}$$

If we square both sides of the expressions in (55) and (56) we can rewrite the inequalities as

$$\left\|\nabla_{\tilde{\boldsymbol{\varphi}}^*} f\left(\mathbf{F}^{(q_2)},\tilde{\boldsymbol{\varphi}}^{(q_2)}\right)-\nabla_{\tilde{\boldsymbol{\varphi}}^*} f\left(\mathbf{F}^{(q_1)},\tilde{\boldsymbol{\varphi}}^{(q_1)}\right)\right\|_2^2\le\left(d^2+de\right)\left\|\mathbf{F}^{(q_2)}-\mathbf{F}^{(q_1)}\right\|_F^2+\left(e^2+de\right)\left\|\tilde{\boldsymbol{\varphi}}^{(q_2)}-\tilde{\boldsymbol{\varphi}}^{(q_1)}\right\|_2^2, \tag{57}$$

$$\left\|\nabla_{\mathbf{F}^*} f\left(\mathbf{F}^{(q_2)},\tilde{\boldsymbol{\varphi}}^{(q_2)}\right)-\nabla_{\mathbf{F}^*} f\left(\mathbf{F}^{(q_1)},\tilde{\boldsymbol{\varphi}}^{(q_1)}\right)\right\|_F^2\le\left(b^2+bc\right)\left\|\mathbf{F}^{(q_2)}-\mathbf{F}^{(q_1)}\right\|_F^2+\left(c^2+bc\right)\left\|\tilde{\boldsymbol{\varphi}}^{(q_2)}-\tilde{\boldsymbol{\varphi}}^{(q_1)}\right\|_2^2, \tag{58}$$

where we applied the relation $2\left\|\mathbf{F}^{(q_2)}-\mathbf{F}^{(q_1)}\right\|_F\left\|\tilde{\boldsymbol{\varphi}}^{(q_2)}-\tilde{\boldsymbol{\varphi}}^{(q_1)}\right\|_2\le\left\|\mathbf{F}^{(q_2)}-\mathbf{F}^{(q_1)}\right\|_F^2+\left\|\tilde{\boldsymbol{\varphi}}^{(q_2)}-\tilde{\boldsymbol{\varphi}}^{(q_1)}\right\|_2^2$.

Combining both expressions and applying the square root to both sides we finally obtain

$$\left(\left\|\nabla_{\mathbf{F}^*} f\left(\mathbf{F}^{(q_2)},\tilde{\boldsymbol{\varphi}}^{(q_2)}\right)-\nabla_{\mathbf{F}^*} f\left(\mathbf{F}^{(q_1)},\tilde{\boldsymbol{\varphi}}^{(q_1)}\right)\right\|_F^2+\left\|\nabla_{\tilde{\boldsymbol{\varphi}}^*} f\left(\mathbf{F}^{(q_2)},\tilde{\boldsymbol{\varphi}}^{(q_2)}\right)-\nabla_{\tilde{\boldsymbol{\varphi}}^*} f\left(\mathbf{F}^{(q_1)},\tilde{\boldsymbol{\varphi}}^{(q_1)}\right)\right\|_2^2\right)^{\frac{1}{2}}$$

$$\leq \left( \left(b^2+bc+d^2+de\right)\left\|\mathbf{F}^{(q_2)}-\mathbf{F}^{(q_1)}\right\|_F^2+\left(c^2+bc+e^2+de\right)\left\|\tilde{\boldsymbol{\varphi}}^{(q_2)}-\tilde{\boldsymbol{\varphi}}^{(q_1)}\right\|_2^2\right)^{\frac{1}{2}}$$

$$\leq L\left(\left\|\mathbf{F}^{(q_2)}-\mathbf{F}^{(q_1)}\right\|_F^2+\left\|\tilde{\boldsymbol{\varphi}}^{(q_2)}-\tilde{\boldsymbol{\varphi}}^{(q_1)}\right\|_2^2\right)^{\frac{1}{2}}. \quad (59)$$

This completes the proof. ■

## APPENDIX C

## PROOF OF THEOREM 2

The following proof is based on the background provided in [25]. First, using definition (6), we can write the update of the monitoring variables (step 6 of Algorithm 1) as

$$\left(\mathbf{U}^{(q+1)},\tilde{\mathbf{v}}^{(q+1)}\right)=\underset{(\mathbf{X},\mathbf{x})}{\operatorname{argmin}}\ \mathcal{I}_{\mathcal{C}}(\mathbf{X})+\operatorname{Re}\left\{\operatorname{Tr}\left\{\left(\nabla_{\mathbf{F}^*}f\left(\mathbf{F}^{(q)},\tilde{\boldsymbol{\varphi}}^{(q)}\right)\right)^H\left(\mathbf{X}-\mathbf{F}^{(q)}\right)\right\}\right\}$$
$$+\frac{1}{2\alpha}\left\|\mathbf{X}-\mathbf{F}^{(q)}\right\|_F^2+\mathcal{I}_{\mathcal{U}_{N_{pan}N_{RIS}}}(\mathbf{x})+\operatorname{Re}\left\{\left(\nabla_{\tilde{\boldsymbol{\varphi}}^*}f\left(\mathbf{F}^{(q)},\tilde{\boldsymbol{\varphi}}^{(q)}\right)\right)^H\left(\mathbf{x}-\tilde{\boldsymbol{\varphi}}^{(q)}\right)\right\}+\frac{1}{2\alpha}\left\|\mathbf{x}-\tilde{\boldsymbol{\varphi}}^{(q)}\right\|_2^2 \quad (60)$$

where we work with the concatenated vectors (46) and $\tilde{\mathbf{v}}=\left[\mathbf{v}_1^{\,T},\ldots,\mathbf{v}_{N_{pan}}^{\,T}\right]^T$. Considering that $\left(\mathbf{F}^{(0)},\tilde{\boldsymbol{\varphi}}^{(0)}\right)$ are initialized over the feasible set then we have $\mathcal{I}_{\mathcal{C}}\left(\mathbf{F}^{(q)}\right)=0$, $\mathcal{I}_{\mathcal{U}_{N_{pan}N_{RIS}}}\left(\tilde{\boldsymbol{\varphi}}^{(q)}\right)=0$, $\mathcal{I}_{\mathcal{C}}\left(\mathbf{U}^{(q+1)}\right)=0$ and $\mathcal{I}_{\mathcal{U}_{N_{pan}N_{RIS}}}\left(\tilde{\mathbf{v}}^{(q+1)}\right)=0$ for any $q$. Furthermore, since $\left(\mathbf{U}^{(q+1)},\tilde{\mathbf{v}}^{(q+1)}\right)$ minimizes (60) we can write

$$\operatorname{Re}\left\{\operatorname{Tr}\left\{\left(\nabla_{\mathbf{F}^*}f\left(\mathbf{F}^{(q)},\tilde{\boldsymbol{\varphi}}^{(q)}\right)\right)^H\left(\mathbf{U}^{(q+1)}-\mathbf{F}^{(q)}\right)\right\}+\left(\nabla_{\tilde{\boldsymbol{\varphi}}^*}f\left(\mathbf{F}^{(q)},\tilde{\boldsymbol{\varphi}}^{(q)}\right)\right)^H\left(\tilde{\mathbf{v}}^{(q+1)}-\tilde{\boldsymbol{\varphi}}^{(q)}\right)\right\}$$
$$\leq -\frac{1}{2\alpha}\left(\left\|\mathbf{U}^{(q+1)}-\mathbf{F}^{(q)}\right\|_F^2+\left\|\tilde{\mathbf{v}}^{(q+1)}-\tilde{\boldsymbol{\varphi}}^{(q)}\right\|_2^2\right) \quad , \quad (61)$$

According to Theorem 1, $f\left(\mathbf{F},\tilde{\boldsymbol{\varphi}}\right)$ has Lipschitz continuous gradients over the feasible set with constant $L$. Therefore, according to the Descent Lemma ([38], chapter 5) we can write the following inequality

$$f\left(\mathbf{U}^{(q+1)},\tilde{\mathbf{v}}^{(q+1)}\right)\leq f\left(\mathbf{F}^{(q)},\tilde{\boldsymbol{\varphi}}^{(q)}\right)+\operatorname{Re}\left\{\operatorname{Tr}\left\{\left(\nabla_{\mathbf{F}^*}f\left(\mathbf{F}^{(q)},\tilde{\boldsymbol{\varphi}}^{(q)}\right)\right)^H\left(\mathbf{U}^{(q+1)}-\mathbf{F}^{(q)}\right)\right\}\right\}+\frac{L}{2}\left\|\mathbf{U}^{(q+1)}-\mathbf{F}^{(q)}\right\|_F^2$$
$$+\operatorname{Re}\left\{\left(\nabla_{\tilde{\boldsymbol{\varphi}}^*}f\left(\mathbf{F}^{(q)},\tilde{\boldsymbol{\varphi}}^{(q)}\right)\right)^H\left(\tilde{\mathbf{v}}^{(q+1)}-\tilde{\boldsymbol{\varphi}}^{(q)}\right)\right\}+\frac{L}{2}\left\|\tilde{\mathbf{v}}^{(q+1)}-\tilde{\boldsymbol{\varphi}}^{(q)}\right\|_2^2. \quad (62)$$

Combining this expression with (61) allows us to write

$$f\left(\mathbf{U}^{(q+1)},\tilde{\mathbf{v}}^{(q+1)}\right)\leq f\left(\mathbf{F}^{(q)},\tilde{\boldsymbol{\varphi}}^{(q)}\right)-\left(\frac{1}{2\alpha}-\frac{L}{2}\right)\left(\left\|\mathbf{U}^{(q+1)}-\mathbf{F}^{(q)}\right\|_F^2+\left\|\tilde{\mathbf{v}}^{(q+1)}-\tilde{\boldsymbol{\varphi}}^{(q)}\right\|_2^2\right). \quad (63)$$

Since we have $\alpha<\frac{1}{L}$ then we obtain $f\left(\mathbf{U}^{(q+1)},\tilde{\mathbf{v}}^{(q+1)}\right)\leq f\left(\mathbf{F}^{(q)},\tilde{\boldsymbol{\varphi}}^{(q)}\right)$. Noting that the update of $\left(\mathbf{F}^{(q+1)},\tilde{\boldsymbol{\varphi}}^{(q+1)}\right)$ is performed according to steps 7-11 of Algorithm 1, then we can conclude that

$$f\left(\mathbf{F}^{(q+1)},\tilde{\boldsymbol{\varphi}}^{(q+1)}\right)\leq f\left(\mathbf{U}^{(q+1)},\tilde{\mathbf{v}}^{(q+1)}\right)\leq f\left(\mathbf{F}^{(q)},\tilde{\boldsymbol{\varphi}}^{(q)}\right), \tag{64}$$

i.e., the algorithm is nonincreasing over the sequence $\left\{\left(\mathbf{F}^{(q)},\tilde{\boldsymbol{\varphi}}^{(q)}\right)\right\}$. Furthermore, sequences $\left\{\left(\mathbf{F}^{(q)},\tilde{\boldsymbol{\varphi}}^{(q)}\right)\right\}$ and $\left\{\left(\mathbf{U}^{(q)},\tilde{\mathbf{v}}^{(q)}\right)\right\}$ are bounded since for finite valued channel matrices $f(\mathbf{F},\tilde{\boldsymbol{\varphi}})$ is bounded from below and the constraint sets $\mathcal{U}_{N_{RIS}}$ and $\mathcal{C}$ are bounded and closed. Therefore, $\left\{\left(\mathbf{F}^{(q)},\tilde{\boldsymbol{\varphi}}^{(q)}\right)\right\}$ has accumulation points, with $f(\mathbf{F},\tilde{\boldsymbol{\varphi}})$ taking the same value at those points as it is nonincreasing. We denote this value as $\hat{f}$. Combining (63) and (64) we have

$$\left(\frac{1}{2\alpha}-\frac{L}{2}\right)\left(\left\|\mathbf{U}^{(q+1)}-\mathbf{F}^{(q)}\right\|_F^2+\left\|\tilde{\mathbf{v}}^{(q+1)}-\tilde{\boldsymbol{\varphi}}^{(q)}\right\|_2^2\right)\leq f\left(\mathbf{F}^{(q)},\tilde{\boldsymbol{\varphi}}^{(q)}\right)-f\left(\mathbf{F}^{(q+1)},\tilde{\boldsymbol{\varphi}}^{(q+1)}\right). \tag{65}$$

Applying a summation over $q=0,\ldots,+\infty$ on both sides yields

$$\left(\frac{1}{2\alpha}-\frac{L}{2}\right)\sum_{q=0}^{+\infty}\left(\left\|\mathbf{U}^{(q+1)}-\mathbf{F}^{(q)}\right\|_F^2+\left\|\tilde{\mathbf{v}}^{(q+1)}-\tilde{\boldsymbol{\varphi}}^{(q)}\right\|_2^2\right)\leq f\left(\mathbf{F}^{(0)},\tilde{\boldsymbol{\varphi}}^{(0)}\right)-\hat{f}<+\infty, \tag{66}$$

which implies

$$\left\|\mathbf{U}^{(q+1)}-\mathbf{F}^{(q)}\right\|_F^2\to 0,\quad \left\|\tilde{\mathbf{v}}^{(q+1)}-\tilde{\boldsymbol{\varphi}}^{(q)}\right\|_2^2\to 0 \tag{67}$$

when $q\to+\infty$. This means that if we denote any accumulation point of $\left\{\left(\mathbf{F}^{(q)},\tilde{\boldsymbol{\varphi}}^{(q)}\right)\right\}$ as $\left\{(\mathbf{F}',\tilde{\boldsymbol{\varphi}}')\right\}$, we can always build a subsequence $\left\{\left(\mathbf{U}^{(q_j)},\tilde{\mathbf{v}}^{(q_j)}\right)\right\}$ that converges to that point when $j\to+\infty$. Furthermore, $f(\mathbf{F},\tilde{\boldsymbol{\varphi}})$ has a continuous gradient and, thus, $f\left(\mathbf{U}^{(q_j+1)},\tilde{\mathbf{v}}^{(q_j+1)}\right)\to f(\mathbf{F}',\tilde{\boldsymbol{\varphi}}')$. $\left(\mathbf{U}^{(q+1)},\tilde{\mathbf{v}}^{(q+1)}\right)$ is computed as in step 6 of Algorithm 1 which, according to (6) allows us to write

$$\begin{cases}0\in\partial\mathcal{I}_{\mathcal{C}}\left(\mathbf{U}^{(q+1)}\right)+\nabla_{\mathbf{F}^*}f\left(\mathbf{F}^{(q)},\tilde{\boldsymbol{\varphi}}^{(q)}\right)+\dfrac{1}{2\alpha}\left(\mathbf{U}^{(q+1)}-\mathbf{F}^{(q)}\right)\\ 0\in\partial\mathcal{I}_{\mathcal{U}_{N_{pan}N_{RIS}}}\left(\tilde{\mathbf{v}}^{(q+1)}\right)+\nabla_{\tilde{\boldsymbol{\varphi}}^*}f\left(\mathbf{F}^{(q)},\tilde{\boldsymbol{\varphi}}^{(q)}\right)+\dfrac{1}{2\alpha}\left(\tilde{\mathbf{v}}^{(q+1)}-\tilde{\boldsymbol{\varphi}}^{(q)}\right)\end{cases}. \tag{68}$$

This condition can be rewritten as

$$0\in\partial G\left(\mathbf{U}^{(q+1)},\tilde{\mathbf{v}}^{(q+1)}\right)-\left(\nabla_{\mathbf{F}^*}f\left(\mathbf{U}^{(q+1)},\tilde{\mathbf{v}}^{(q+1)}\right)-\nabla_{\mathbf{F}^*}f\left(\mathbf{F}^{(q)},\tilde{\boldsymbol{\varphi}}^{(q)}\right),\nabla_{\tilde{\boldsymbol{\varphi}}^*}f\left(\mathbf{U}^{(q+1)},\tilde{\mathbf{v}}^{(q+1)}\right)-\nabla_{\tilde{\boldsymbol{\varphi}}^*}f\left(\mathbf{F}^{(q)},\tilde{\boldsymbol{\varphi}}^{(q)}\right)\right)$$
$$+\frac{1}{2\alpha}\left(\mathbf{U}^{(q+1)}-\mathbf{F}^{(q)},\tilde{\mathbf{v}}^{(q+1)}-\tilde{\boldsymbol{\varphi}}^{(q)}\right) \tag{69}$$

with

$$G\left(\mathbf{U}^{(q+1)},\tilde{\mathbf{v}}^{(q+1)}\right)=\mathcal{I}_{\mathcal{C}}\left(\mathbf{U}^{(q+1)}\right)+\partial\mathcal{I}_{\mathcal{U}_{N_{pan}N_{RIS}}}\left(\tilde{\mathbf{v}}^{(q+1)}\right)+f\left(\mathbf{U}^{(q+1)},\tilde{\mathbf{v}}^{(q+1)}\right) \tag{70}$$

and

$$\partial G\left(\mathbf{U}^{(q+1)},\tilde{\mathbf{v}}^{(q+1)}\right)=\left(\partial\mathcal{I}_{\mathcal{C}}\left(\mathbf{U}^{(q+1)}\right)+\nabla_{\mathbf{F}^*}f\left(\mathbf{U}^{(q+1)},\tilde{\mathbf{v}}^{(q+1)}\right),\partial\mathcal{I}_{\mathcal{U}_{N_{pan}N_{RIS}}}\left(\tilde{\mathbf{v}}^{(q+1)}\right)+\nabla_{\tilde{\boldsymbol{\varphi}}^*}f\left(\mathbf{U}^{(q+1)},\tilde{\mathbf{v}}^{(q+1)}\right)\right). \tag{71}$$

To satisfy condition (69) we must have

$$\left(\nabla_{\mathbf{F}^*} f\left(\mathbf{U}^{(q+1)},\tilde{\mathbf{v}}^{(q+1)}\right)-\nabla_{\mathbf{F}^*} f\left(\mathbf{F}^{(q)},\tilde{\boldsymbol{\varphi}}^{(q)}\right),\nabla_{\tilde{\boldsymbol{\varphi}}^*} f\left(\mathbf{U}^{(q+1)},\tilde{\mathbf{v}}^{(q+1)}\right)-\nabla_{\tilde{\boldsymbol{\varphi}}^*} f\left(\mathbf{F}^{(q)},\tilde{\boldsymbol{\varphi}}^{(q)}\right)\right) -\frac{1}{2\alpha}\left(\mathbf{U}^{(q+1)}-\mathbf{F}^{(q)},\tilde{\mathbf{v}}^{(q+1)}-\tilde{\boldsymbol{\varphi}}^{(q)}\right)\in\partial G\left(\mathbf{U}^{(q+1)},\tilde{\mathbf{v}}^{(q+1)}\right) \tag{72}$$

Working with the subsequence $\left(\mathbf{U}^{(q_j+1)},\tilde{\mathbf{v}}^{(q_j+1)}\right)\rightarrow\left(\mathbf{F}',\tilde{\boldsymbol{\varphi}}'\right)$, and since $f\left(\mathbf{U}^{(q_j+1)},\tilde{\mathbf{v}}^{(q_j+1)}\right)\rightarrow f\left(\mathbf{F}',\tilde{\boldsymbol{\varphi}}'\right)$, then the subgradient in the left side of (72) must tend to a subgradient in $\partial G\left(\mathbf{F}',\tilde{\boldsymbol{\varphi}}'\right)$ when $j\rightarrow+\infty$. Furthermore, we can express its as

$$\begin{aligned}&\left(\left\|\nabla_{\mathbf{F}^*} f\left(\mathbf{U}^{(q_j+1)},\tilde{\mathbf{v}}^{(q_j+1)}\right)-\nabla_{\mathbf{F}^*} f\left(\mathbf{F}^{(q_j)},\tilde{\boldsymbol{\varphi}}^{(q_j)}\right)-\frac{1}{2\alpha}\left(\mathbf{U}^{(q_j+1)}-\mathbf{F}^{(q_j)}\right)\right\|_F^2\right.\\&\left.+\left\|\nabla_{\tilde{\boldsymbol{\varphi}}^*} f\left(\mathbf{U}^{(q_j+1)},\tilde{\mathbf{v}}^{(q_j+1)}\right)-\nabla_{\tilde{\boldsymbol{\varphi}}^*} f\left(\mathbf{F}^{(q_j)},\tilde{\boldsymbol{\varphi}}^{(q_j)}\right)-\frac{1}{2\alpha}\left(\tilde{\mathbf{v}}^{(q_j+1)}-\tilde{\boldsymbol{\varphi}}^{(q_j)}\right)\right\|_2^2\right)^{\frac{1}{2}}\\&\leq\left(L+\frac{1}{2\alpha}\right)\sqrt{\left\|\mathbf{U}^{(q_j+1)}-\mathbf{F}^{(q_j)}\right\|_F^2+\left\|\tilde{\mathbf{v}}^{(q_j+1)}-\tilde{\boldsymbol{\varphi}}^{(q_j)}\right\|_2^2}.\end{aligned} \tag{73}$$

Using (67) we can then conclude that this norm tends to 0 and thus $0\in\partial G\left(\mathbf{F}',\tilde{\boldsymbol{\varphi}}'\right)$, i.e., $\left(\mathbf{F}',\tilde{\boldsymbol{\varphi}}'\right)$ is a critical point of (4). ■